# Optical Observations Reveal Strong Evidence for High Energy Neutrino Progenitor

V.M. Lipunov[1,2], V.G. Kornilov[1,2], K.Zhirkov[1], E. Gorbovskoy[2], N.M. Budnev[4], D.A.H.Buckley[3], R. Rebolo[5], M. Serra-Ricart[5], R. Podesta[9,10], N .Tyurina[2], O. Gress[4,2], Yu.Sergienko[8], V. Yurkov[8], A. Gabovich[8], P.Balanutsa[2], I.Gorbunov[2], D.Vlasenko[1,2], F.Balakin[1,2], V.Topolev[1], A.Pozdnyakov[1], A.Kuznetsov[2], V.Vladimirov[2], A. Chasovnikov[1], D. Kuvshinov[1,2], V.Grinshpun[1,2], E.Minkina[1,2], V.B.Petkov[7], S.I.Svertilov[2,6], C. Lopez[9], F. Podesta[9], H.Levato[10], A. Tlatov[11], B. Van Soelen[12], S. Razzaque[13], M. Böttcher[14]

[1] M.V.Lomonosov Moscow State University, Physics Department, Leninskie gory, GSP-1, Moscow, 119991, Russia;
lipunov2007@gmail.com
[2] M.V.Lomonosov Moscow State University, SAI, Universitetsky pr., 13, Moscow, 119234, Russia;
[3] South African Astrophysical Observatory, PO Box 9, 7935 Observatory, Cape Town, South Africa;
[4] Irkutsk State University, Applied Physics Institute, 20, Gagarin blvd, 664003, Irkutsk, Russia;
[5] Instituto de Astrofisica de Canarias Via Lactea, s/n E38205 - La Laguna (Tenerife), Spain;
[6] Lomonosov Moscow State University, Skobeltsyn Institute of Nuclear Physics, Moscow 119234, Russia;
[7] Institute for Nuclear Research of RAS, Moscow 117312, Russia;
[8] Blagoveschensk State Pedagogical University, Lenin str., 104, Blagoveschensk 675000, Russia;
[9] Observatorio Astronomico Felix Aguilar(OAFA), Avda Benavides s/n, Rivadavia, El Leonsito, Argentina;
[10] San Juan National University, Casilla de Correo 49, 5400 San Juan, Argentina;
[11] Kislovodsk Solar Station of the Pulkovo Observatory RAS, P.O.Box 45, ul. Gagarina 100, Kislovodsk 357700, Russia
[12] Physics Dept., Department of Physics, University of the Free State, PO Box 339, Bloemfontein 9300, South Africa
[13] Department of Physics, University of Johannesburg, PO Box 559, Auckland Park, South Africa
[14] Centre for Space Research, North-West University, Potchefstroom 2520, South Africa

**Abstract**

We present the earliest astronomical observation of a high energy neutrino error box in which its variability was discovered after high-energy neutrinos detection. The one robotic telescope of the MASTER global international network (Lipunov et al. 2010) automatically imaged the error box of the very high-energy neutrino event IceCube-170922A. Observations were carried out in minute after the IceCube-170922A neutrino event was detected by the IceCube observatory at the South Pole. MASTER found the blazar TXS 0506+056 to be in the off-state after one minute and then switched to the on-state no later than two hours after the event. The effect is observed at a 50-sigma significance level. Also we present own unique 16-years light curve of blazar TXS 0506+056 (518 data set).

## 1. Introduction

It is still not possible to understand where cosmic neutrinos of high energies come from and astronomical robots have joined in solving this problem. A few years ago, the MASTER[15] Global Network of Robot Telescopes (Lipunov et al. 2010) began to respond to IceCube alerts. On 22d of September, 2017 the robotic telescope of the MASTER global network automatically imaged the error box of the high-energy neutrino event IceCube-170922A (Kopper et al. 2017). Observations were carried out 27 s after receiving the alert, i.e., 73 s after the IceCube-170922A neutrino event was detected by the IceCube observatory at the South Pole (Lipunov et al. 2018a). However, surprising details of these observations are published only now. We recently recalibrated these images using the Gaia (Brown et al. 2018) catalog as the source of reference stars, and found the BL Lacertae type blazar TXS 0506+056 (IceCube et al. 2018ab) to be in the off-state after one minute and then switched to the on-state no later than two hours after the event. The effect is observed at Δm =0.790+-0.016 (a 50-σ significance level).

Identification of the astrophysical sources of very high energy neutrinos still remains one of the most exiting enigmas of the Universe. It was with the aim of detecting sources of VHE neutrinos that unique observatories were built at the South Pole (IceCube et al. 2017,2018ab), in the Mediterranean Sea (ANTARES, Ageron et al. 2011), in the deepest Baikal lake (Balkanov et al. 2002), and under the Caucasus mountain ridge (Baksan Neutrino Observatory, Boliev et al. 2018).

---

[15] http://observ.pereplet.ru

Unlike very high energy cosmic rays, electrically neutral neutrinos freely propagate across the Universe undeflected by the intergalactic magnetic field and unattenuated by the interaction with cosmic background emission (IceCube et al.2018ab). Hence neutrino trajectories point to the sources of these particles.

Unfortunately, scattering of light in ice or in water, the working media of sparsely instrumented neutrino detectors, blur the positional error regions, whose sizes are currently comparable to one square degree. Therefore finding a blazar within the error box of a VHE neutrino event cannot be considered sufficient to prove that blazars are actually progenitors of these particles. Detecting some non-standard event from the supposed source at a time close to the neutrino event is required. For example, a blazar emitting gamma and cosmic rays and showing a sharp flux variation near the neutrino detection time would provide compelling evidence of the association of the neutrino event with a known astrophysical object. The first candidate object for an astrophysical neutrino event was the blazar TXS 0506+056 (IceCube et al. 2018b) found inside the error box of the IceCube-170922A neutrino event. This blazar turned out to be located at a distance of ~3.7 billion light years (its redshift is z = 0.3365+/- 0.0010) (Paiano et al. 2018).

Although the blazar was in the gamma-ray active state, this state started several months before the neutrino event. Detection of high-energy particles (175Gev) began one week after, and the optical, x-ray, and gamma-ray emission was observed with low temporal resolution and showed no appreciable variations near the detection time (IceCube et al. 2018b).

Therefore although when combining the available data suggested that TXS 0506+056 was a very promising high energy neutrino source optical candidate, the temporal resolution of multi-messenger data did not provide conclusive evidence at the time and the object remained just a likely, but still debatable, candidate (IceCube et al. 2018b). In this Letter we report conclusive detection of light variation of the blazar TXS 0506+056 just several minutes after the neutrino event, which ended no later than after two hours. For comparison, nearest ASAS-SN, Kiso/KWFC and Kanata/HONIR optical observations do not show the same decrease in optical brightness (IceCube et al. 2018b) because they started 18 hours after MASTER observations when the effect disappeared.

## 2. MASTER real-time optical observations of the IceCube-170922A error box

As the leader of early gamma-ray burst observations, MASTER Global Robotic Net has an almost 20-year long experience with real-time rapid pointings to gamma-ray burst alerts within the first minute of the alert *(*Troja et al. 2017; Ershova et al. 2020). Starting with 2015, MASTER Global Robotic Net has been actively participating in the program of fast optical support of major physical and astrophysical experiments, such as detection of very high energy neutrinos (ANTARES (Dornic et al. 2015; Gress et al. 2019), IceCube (Aartsen et al. 2017) , Baksan (Lipunov et al. 2019a), gravitational waves (LIGO/VIRGO collaboration (Abbott et al. 2016,2017, Lipunov et al. 2017ab)), and Fast Radio Bursts (FRB (Lipunov et al. 2018b)). The favorable arrangement of MASTER sites makes it possible to inspect all gravitational-wave error boxes. MASTER made the crucial contribution to the optical support of the first gravitational-wave event GW 150914 by inspecting the largest part of the error box *(*Abbott et al. 2016, Lipunov et al. 2017b*)*. On 17[th] of August, 2017 MASTER, together with five other telescopes, performed the first ever optical localization of a gravitational-wave source by acquiring early optical images of the Kilonova from GW 170817 *(*Abbot et al. 2017, Lipunov et al. 2017a*)*.

It was the successful finding of the kilonova (Abbot et al. 2017) at the location of a neutron-star collision for the GW 170817 event that made the MASTER team focus on analyzing this extremely important event, until the end of 2017. That is why we only published the results of our observations of the IceCube-170922 in 2018 (Lipunov et al. 2018a) and not in September, 2017. However, our robotic telescopes made everything themselves. This article was attended by eight sites of the MASTER Global Robotic Network: MASTER-Amur (Blagoveshchensk, Russian Far East), MASTER-Tunka (located near Baykal Lake in Siberia, Russia), MASTER-Vostryakovo (near the Moscow), MASTER-Tavrida (Crimea, Russia), MASTER-Kislovodsk (Caucasus, Russia), MASTER-SAAO (South Africa), MASTER-IAC (Tenerife, Spain), and MASTER-OAFA (Argentina).

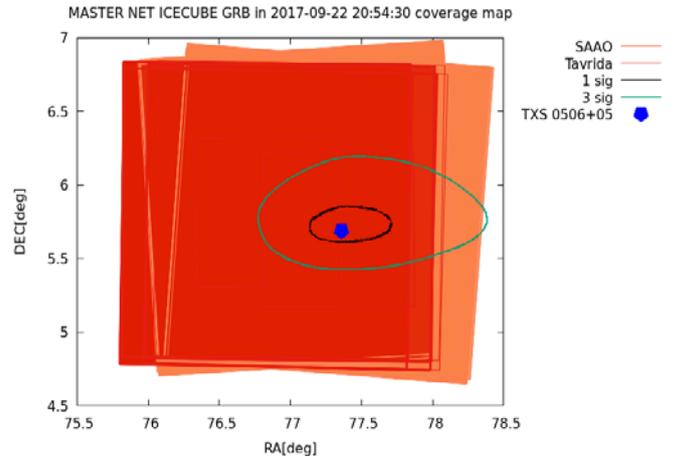

So on 22d of September, 2017, after receiving an alert from IceCube 40 seconds after the neutrino event, MASTER-Tavrida telescope acquired the first three images, starting from 2017-09-22 20:55:43 UT. The field of view of the MASTER telescope has a size of four square degrees (Lipunov et al. 2010) and fully covers the final field of view of IceCube (Figure 1, Brown et al. 2018)

**Figure 1.** The fields of view of our MASTER-Tavrida and MASTER-SAAO telescopes is shown in red and orange in translucent color. MASTER nearly fully covers of the final IceCube-170922 error box during first hours observations. The black and green ovals correspond to the 50% and 90% probability levels, respectively. The blue dot shows the location of the TXS 0506+056 blazar.

Despite the large zenith distance (84 degrees), our 180-second frames reached a limiting magnitude of 19.0m. Hence the TXS 0506+056 blazar at the time of the alert was a 15.12 +- 0.01 magnitude object in all three frames acquired over ~15 minutes (the light curve is shown in Figure 2). There was the faintest blazar brightness over the full period of this alert.

After 2 hours, at 2017-09-22 23:11:36 UT, the flux from the blazar increased in brightness by a factor of two and reached $14.33^m \pm 0.01^m$. We emphasize that here in the text we give the averaged values for the triples of frames in the first minutes and after two hours (compare Table 1). Hence our observations show at an extremely high confidence level of 50-$\sigma$ that within several minutes of the neutrino event, the blazar was in an anomalously extinguished state.

This conclusion was fully confirmed during the two days after the alert, when the MASTER-SAAO robotic telescope joined the blazar observation campaign (Lipunov et al. 2018a)

Was it a unique event for the **TXS 0506+056** blazar**?**

### 3.MASTER optical history of TXS 0506+056.

There are 518 2x2 square-degrees images in the MASTER Global Robotic Net database, starting from 2005, when we had only one MASTER I telescope located near Moscow (Lipunov et al. 2010, MASTER-Vostryakovo in Table). All MASTER telescopes have identical equipment, which gives us the ability to make photometry in one system (Lipunov et al. 2010, 2019b).

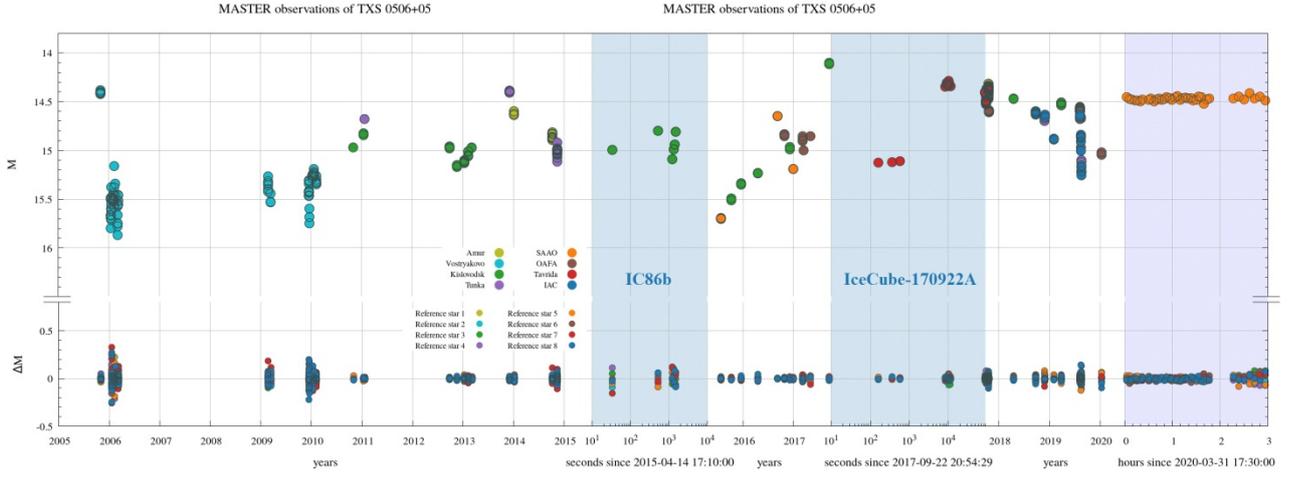

**Figure 2**. MASTER light curve of the TXS 0506+056 blazar for the 16 years. Archive light curve of the TXS 0506+056 blazar based on observations made by MASTER Global Net robotic telescope of from 2005 until now (red point). Below we see the photometry of 8 reference stars, to the pink and blue panels represent three very narrow episodes in time. The first of these is April 2015 when IceCube IC86b saw a 3.5 sigma excess of the neutrino flux over the background (IceCube 2018a) . The second is the 22 September 2017 event (IceCube et al.2018b). Logarithmic time is shown in seconds from the neutrino trigger. It is easy to easily see the rapid change in the luminosity of the blazar in ~2 times. Finally, the third episode is a uniform blazar monitoring timeline in the first quarter of 2020. With these new observations, the total number of observations submitted reached 518 (See the photometry table).

Figure 2 shows our photometry of the blazar over the last 16 years. We chose 8 Gaia catalog stars, having the brightness and color similar to those of the blazar, as photometric reference stars, and estimated the errors of individual photometric measurements from the scatter of the magnitudes of these reference stars. We also checked these stars for rapid and long-term variability and found them to be quite stable. We found three times when the brightness of the blazar varied ~0.5$^m$ more than 1-3 significance level. The first such time was in 2006, when IceCube neutrino observatory was not yet operating. The second time was in April 2015 (substantial increase of neutrino signal IC86b). The MASTER observation date is in April but statistically close to Gaussian IceCube half year window 9/2014 to 3/2015 (IceCube 2018a). And the third time was in September 2017, when the **IceCube-170922 (IC86C)** event occurred (IceCube et al. 2018b)**.**

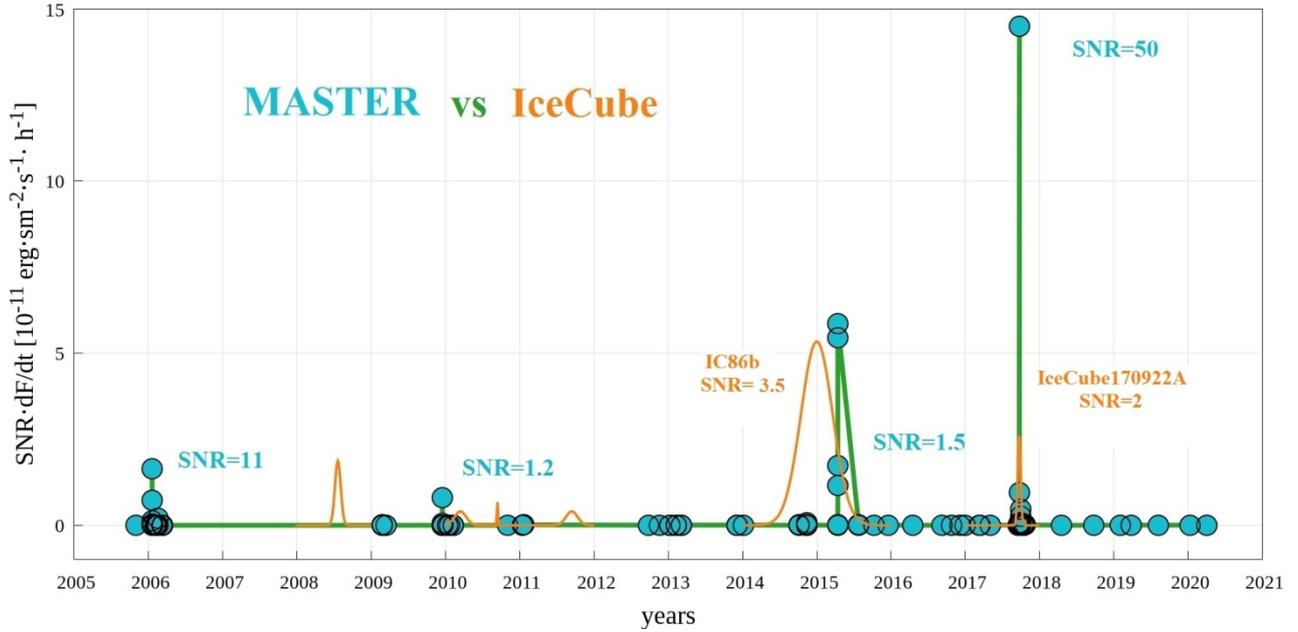

**Figure 3.** TXS 0506 + 056 optical blazar variability rate history. Flux deriviation the multiplied by the signal-to-noise ratio (blue). The orange curves schematically show the Gaussian analysis of the archive data of IceCube (2018a).

The special interest is not even the flux variations themselves, but the rate of their change. To illustrate this, we built a history of the rate of change of the optical flux (Figure 3). To remove the noise we averaged the close points. Differential flux multiplied by the signal-to-noise ratio Y:

$$Y = SNR \cdot (F_i - F_{i-1})/(t_i - t_{i-1}),$$

where SNR is the $(F_i - F_{i-1})/s$, $F_i$ and s – is the optical flux and error of two adjacent measurements at the mid-exposure time $t_i$.

The event of September 22, 2017 has outstanding characteristics in terms of flux derivation and signal-to-noise ratio. Recently, we conducted detailed blazar monitoring over several nights at the right end of Figure 3. As we see in the usual state, the blazar is stable at times of several hours and even days with an accuracy of ~ 0.02 mag. This means that the instability we discovered on 22 September 2017, a few minutes after the neutrino alert, has a reliability of about 40 sigma, and by this criterion. This result shows the power of fast alert observations of sources of ultrahigh energy particles.

### 4. Discussion.

We find for the adopted set of cosmological parameters (Ade et al. 2016) $H_0 = 67.8$ km/s/Mpc (the Hubble constant), $\Omega_m = 0.308$, $\Omega_\Lambda = 0.692$ (the matter and vacuum density) that several minutes after the neutrino event the optical isotropic luminosity of the blazar was $L_{opt}$ ~$4.3 \cdot 10^{45}$ erg/s and after two hours it returned to the typical level within several weeks of the neutrino event, ~$9.7 \cdot 10^{45}$ erg/s (we include galactic absoption $A_B$=0.4 (Schlegel et al. 1998)).

The generally accepted picture is that blazar radiation arises from a relativistic jet directed toward us. The boosted jet gamma factor is moderate Γ ~ 10. In the shock wave at the front of the jet there is an acceleration of protons to ultrahigh energies, which in turn collide with target photons and generate pion production. The decay of pions, in turn, gives rise to a muon neutrino that registers an IceCube detection and high gamma photons detected by the Fermi gamma-ray observatory.

During the period within ~2 weeks around the neutrino event detection time the 0.1 – 100 Gev gamma-ray luminosity was 1.3·10^47 erg/s (IceCube et al 2018b). Note that the neutrino luminosity of the quasar was equal to about *L_v ≈ 4 ·10^47 erg/s,* which is appreciably higher and, evidently, closer to the gamma-ray luminosity. However, this is not surprising because neutrinos and gamma-ray emission have the same source of energy --- that of high-energy protons accelerated by the central supermassive black hole. Two branches of reactions

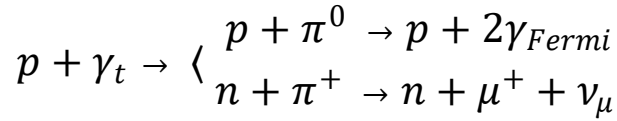

$$p + \gamma_t \to \begin{cases} p + \pi^0 \to p + 2\gamma_{Fermi} \\ n + \pi^+ \to n + \mu^+ + \nu_\mu \end{cases}$$

produce neutral and charged pions, which then decay into and gamma-ray photons ($\gamma_{Fermi}$) and muon neutrinos ($\nu_\mu$). Both pion-birth reactions are of the threshold type *(Hayakava, Yamamoto 1963)*. Let us assume that $\pi$ mesons are born as a result of the collision of relativistic-jet protons with target photons $\gamma_t$. Note that neutrinos and gamma-ray photons carry away several percent of the proton energy. Hence the neutrino luminosity of blazars is determined just by the birth rate of charged $\pi^+$ mesons. The optical luminosity is somehow related to the neutrino flux. Suppose that these same protons accelerated at the front of a shock wave of a relativistic booster are a source of synchrotron optical emission (Paliya et al. 2020). We simply noticed that optical radiation can also produced as synchrotron radiation of protons in a zone with a reduced magnetic field.

Then it should be expected that with an increase in the neutrino flux, due to the disappearance of protons in proton-photon reactions, the optical synchrotron photons of the protons will drop. The maximal amplitude of the decrease in optical luminosity can be as much as 2 times, since the branches of the reaction proceed with approximately the same probability. This is what we are observing.

How does so much variability arise at large distances? Where do we expect the process of acceleration of protons and nuclei to high energies? In the recent work (O' Riordan et al. 2017) of it is shown that the gamma factor of the jet can vary significantly in minutes due to turbulence in magnetized plasma flow near the horizon of a black hole.

We have yet to learn from where the target photons come (Paliya et al. 2020).

### 5.Conclusion

The event that we discovered, namely the decrease of the brightness of the TXS 0506+056 blazar near the neutrino detection time, provides complementary and very compelling evidence for the link between the blazar and the IceCube-170922 neutrino event. We analyzed archival data (MASTER unique 518 photometry data for 16 years), which we found to be consistent with this fact. We also propose a hypothesis explaining the anticorrelation of the optical and neutrino flux. An increase in neutrino flux means that up to half of the protons disappear. If we assume that these protons produce synchrotron optical radiation, then any increase in neutrino luminosity will lead to a decrease in the optical brightness of the blazar.

We acknowledge support by the Development program of Lomonosov Moscow State University (MASTER equipment). D.A.H.B. acknowledges research support from the South African National Research Foundation. We are grateful to Dr. N.N.Kalmykov and Dr. Yu.Yu.Kovalev for discussion on the proton-photon processes and the Blazar jets orientation.

### ORCID iDs

V.M. Lipunov https://orcid.org/0000-0003-3668-1314
D. A. H. Buckley https://orcid.org/0000-0002-7004-9956
R. Rebolo https://orcid.org/0000-0003-3767-7085
N. M. Budnev https://orcid.org/0000-0002-2104-6687

**Table 1.**
MASTER-NET photometry of the blazar TXS 0506+056  since 2005 till 2020 years in Clear band.

| Julian date | Magnitude (Clear band) | Error of mag. | Site | ref1 | ref2 | ref3 | ref4 | ref5 | ref6 | ref7 | ref8 |
|---|---|---|---|---|---|---|---|---|---|---|---|
| 2453676.51582 | 14.43 | 0.01 | MASTER-Vostryakovo | 14.43 | 14.92 | 14.89 | 14.80 | 14.89 | 14.62 | 15.47 | 15.18 |
| 2453676.51697 | 14.40 | 0.01 | MASTER-Vostryakovo | 14.42 | 14.92 | 14.88 | 14.82 | 14.90 | 14.64 | 15.44 | 15.17 |
| 2453676.51811 | 14.41 | 0.01 | MASTER-Vostryakovo | 14.42 | 14.92 | 14.90 | 14.82 | 14.88 | 14.62 | 15.46 | 15.18 |
| 2453676.60042 | 14.41 | 0.03 | MASTER-Vostryakovo | 14.44 | 14.93 | 14.84 | 14.88 | 14.89 | 14.63 | 15.42 | 15.21 |
| 2453676.60159 | 14.38 | 0.01 | MASTER-Vostryakovo | 14.43 | 14.91 | 14.91 | 14.83 | 14.89 | 14.61 | 15.42 | 15.17 |
| 2453676.60281 | 14.42 | 0.02 | MASTER-Vostryakovo | 14.45 | 14.93 | 14.86 | 14.84 | 14.84 | 14.60 | 15.43 | 15.16 |
| 2453753.35019 | 15.50 | 0.06 | MASTER-Vostryakovo | 14.45 | 14.88 | 14.96 | 14.75 | 14.92 | 14.65 | 15.37 | 15.10 |
| 2453753.35505 | 15.67 | 0.15 | MASTER-Vostryakovo | 14.48 | 14.85 | 15.09 | 14.82 | 14.82 | 14.53 | 15.19 | 15.42 |
| 2453753.35604 | 15.49 | 0.09 | MASTER-Vostryakovo | 14.52 | 14.99 | 14.90 | 14.87 | 14.94 | 14.47 | 15.36 | 15.06 |
| 2453753.38351 | 15.80 | 0.14 | MASTER-Vostryakovo | 14.38 | 14.87 | 14.83 | 14.76 | 14.94 | 14.87 | 15.53 | 14.93 |
| 2453753.38644 | 15.66 | 0.17 | MASTER-Vostryakovo | 14.40 | 14.97 | 14.72 | 14.82 | 14.77 | 14.54 | 15.77 | 15.45 |
| 2453753.38842 | 15.69 | 0.08 | MASTER-Vostryakovo | 14.35 | 14.98 | 14.77 | 14.97 | 14.90 | 14.66 | 15.43 | 15.11 |
| 2453754.30800 | 15.60 | 0.05 | MASTER-Vostryakovo | 14.35 | 14.88 | 14.85 | 14.88 | 14.98 | 14.58 | 15.47 | 15.18 |
| 2453754.30908 | 15.62 | 0.08 | MASTER-Vostryakovo | 14.43 | 14.84 | 14.75 | 14.91 | 14.99 | 14.67 | 15.39 | 15.21 |
| 2453754.31013 | 15.38 | 0.08 | MASTER-Vostryakovo | 14.51 | 15.07 | 14.88 | 14.75 | 14.79 | 14.62 | 15.35 | 15.14 |
| 2453754.34060 | 15.55 | 0.08 | MASTER-Vostryakovo | 14.49 | 15.06 | 14.95 | 14.80 | 14.80 | 14.63 | 15.33 | 15.11 |
| 2453754.34353 | 15.60 | 0.04 | MASTER-Vostryakovo | 14.45 | 14.92 | 14.90 | 14.88 | 14.90 | 14.58 | 15.36 | 15.20 |
| 2453754.34455 | 15.71 | 0.10 | MASTER-Vostryakovo | 14.36 | 14.77 | 14.93 | 14.77 | 14.85 | 14.64 | 15.64 | 15.24 |
| 2453758.29468 | 15.48 | 0.10 | MASTER-Vostryakovo | 14.41 | 14.90 | 14.96 | 14.74 | 14.78 | 14.62 | 15.67 | 15.21 |
| 2453769.27536 | 15.51 | 0.04 | MASTER-Vostryakovo | 14.43 | 14.87 | 14.91 | 14.80 | 14.86 | 14.67 | 15.44 | 15.13 |
| 2453769.27631 | 15.44 | 0.04 | MASTER-Vostryakovo | 14.41 | 14.94 | 14.90 | 14.88 | 14.92 | 14.57 | 15.36 | 15.20 |
| 2453769.27728 | 15.49 | 0.03 | MASTER-Vostryakovo | 14.46 | 14.90 | 14.82 | 14.83 | 14.88 | 14.63 | 15.46 | 15.21 |
| 2453769.29971 | 15.50 | 0.03 | MASTER-Vostryakovo | 14.42 | 14.94 | 14.89 | 14.77 | 14.89 | 14.66 | 15.42 | 15.19 |
| 2453769.30066 | 15.46 | 0.04 | MASTER-Vostryakovo | 14.45 | 14.92 | 14.86 | 14.84 | 14.89 | 14.61 | 15.56 | 15.16 |
| 2453769.30159 | 15.52 | 0.03 | MASTER-Vostryakovo | 14.42 | 14.97 | 14.88 | 14.87 | 14.86 | 14.60 | 15.41 | 15.18 |
| 2453775.25978 | 15.38 | 0.10 | MASTER-Vostryakovo | 14.49 | 15.01 | 14.87 | 14.62 | 15.05 | 14.59 | 15.40 | 15.15 |
| 2453775.26161 | 15.55 | 0.11 | MASTER-Vostryakovo | 14.65 | 14.95 | 14.83 | 14.79 | 14.70 | 14.66 | 15.36 | 15.15 |
| 2453776.24277 | 15.58 | 0.02 | MASTER-Vostryakovo | 14.42 | 14.94 | 14.87 | 14.82 | 14.90 | 14.60 | 15.48 | 15.18 |
| 2453776.24358 | 15.52 | 0.04 | MASTER-Vostryakovo | 14.41 | 14.91 | 14.88 | 14.87 | 14.87 | 14.65 | 15.36 | 15.21 |
| 2453776.27821 | 15.16 | 0.02 | MASTER-Vostryakovo | 14.46 | 14.91 | 14.88 | 14.81 | 14.88 | 14.61 | 15.48 | 15.14 |
| 2453782.29183 | 15.34 | 0.06 | MASTER-Vostryakovo | 14.56 | 14.91 | 14.87 | 14.76 | 14.78 | 14.66 | 15.41 | 15.19 |
| 2453782.31780 | 15.46 | 0.06 | MASTER-Vostryakovo | 14.42 | 14.92 | 14.90 | 14.81 | 14.80 | 14.62 | 15.47 | 15.31 |
| 2453801.22751 | 15.87 | 0.04 | MASTER-Vostryakovo | 14.43 | 14.97 | 14.83 | 14.87 | 14.92 | 14.61 | 15.38 | 15.15 |
| 2453801.22851 | 15.75 | 0.03 | MASTER-Vostryakovo | 14.43 | 14.96 | 14.89 | 14.77 | 14.85 | 14.64 | 15.44 | 15.20 |
| 2453801.22940 | 15.78 | 0.03 | MASTER-Vostryakovo | 14.44 | 14.88 | 14.87 | 14.81 | 14.91 | 14.62 | 15.52 | 15.15 |
| 2453801.25681 | 15.75 | 0.02 | MASTER-Vostryakovo | 14.45 | 14.88 | 14.85 | 14.84 | 14.92 | 14.62 | 15.43 | 15.17 |
| 2453801.25770 | 15.66 | 0.05 | MASTER-Vostryakovo | 14.43 | 14.98 | 14.91 | 14.81 | 14.86 | 14.64 | 15.33 | 15.19 |
| 2453801.25876 | 15.57 | 0.06 | MASTER-Vostryakovo | 14.42 | 14.85 | 14.91 | 14.89 | 14.84 | 14.59 | 15.56 | 15.21 |
| 2453804.23730 | 15.56 | 0.04 | MASTER-Vostryakovo | 14.43 | 14.97 | 14.82 | 14.79 | 14.89 | 14.66 | 15.39 | 15.22 |
| 2453804.23826 | 15.52 | 0.05 | MASTER-Vostryakovo | 14.43 | 14.93 | 14.97 | 14.80 | 14.86 | 14.64 | 15.37 | 15.16 |
| 2453804.23920 | 15.46 | 0.02 | MASTER-Vostryakovo | 14.44 | 14.92 | 14.87 | 14.82 | 14.89 | 14.59 | 15.45 | 15.21 |
| 2453804.26268 | 15.46 | 0.05 | MASTER-Vostryakovo | 14.41 | 14.92 | 14.80 | 14.91 | 14.91 | 14.63 | 15.49 | 15.12 |
| 2453804.26387 | 15.55 | 0.06 | MASTER-Vostryakovo | 14.45 | 14.85 | 14.85 | 14.80 | 14.88 | 14.62 | 15.57 | 15.22 |
| 2453804.26477 | 15.52 | 0.04 | MASTER-Vostryakovo | 14.43 | 14.94 | 14.96 | 14.87 | 14.86 | 14.58 | 15.39 | 15.14 |
| 2454886.25126 | 15.36 | 0.05 | MASTER-Vostryakovo | 14.43 | 14.94 | 14.83 | 14.83 | 14.80 | 14.69 | 15.42 | 15.24 |
| 2454886.25210 | 15.42 | 0.06 | MASTER-Vostryakovo | 14.44 | 14.88 | 14.95 | 14.90 | 14.93 | 14.56 | 15.36 | 15.12 |
| 2454886.25294 | 15.32 | 0.04 | MASTER-Vostryakovo | 14.44 | 14.90 | 14.89 | 14.78 | 14.89 | 14.66 | 15.36 | 15.22 |
| 2454886.27548 | 15.34 | 0.04 | MASTER-Vostryakovo | 14.45 | 14.91 | 14.90 | 14.87 | 14.82 | 14.64 | 15.47 | 15.10 |
| 2454886.27632 | 15.27 | 0.04 | MASTER-Vostryakovo | 14.40 | 14.93 | 14.90 | 14.82 | 14.94 | 14.57 | 15.41 | 15.22 |
| 2454886.27799 | 15.40 | 0.08 | MASTER-Vostryakovo | 14.43 | 14.96 | 14.78 | 14.79 | 14.91 | 14.59 | 15.63 | 15.17 |
| 2454905.26363 | 15.44 | 0.06 | MASTER-Vostryakovo | 14.41 | 14.91 | 14.88 | 14.80 | 14.87 | 14.67 | 15.56 | 15.10 |
| 2454905.26447 | 15.53 | 0.05 | MASTER-Vostryakovo | 14.39 | 14.89 | 14.89 | 14.87 | 14.96 | 14.57 | 15.38 | 15.26 |
| 2454905.26774 | 15.53 | 0.04 | MASTER-Vostryakovo | 14.50 | 14.96 | 14.86 | 14.82 | 14.81 | 14.62 | 15.38 | 15.18 |
| 2455180.49664 | 15.42 | 0.07 | MASTER-Vostryakovo | 14.45 | 14.85 | 14.92 | 14.86 | 14.88 | 14.59 | 15.31 | 15.28 |
| 2455180.49754 | 15.47 | 0.04 | MASTER-Vostryakovo | 14.42 | 14.98 | 14.87 | 14.84 | 14.88 | 14.66 | 15.42 | 15.09 |
| 2455180.49854 | 15.47 | 0.06 | MASTER-Vostryakovo | 14.43 | 14.93 | 14.84 | 14.80 | 14.89 | 14.61 | 15.59 | 15.17 |
| 2455180.54417 | 15.43 | 0.10 | MASTER-Vostryakovo | 14.50 | 14.96 | 14.91 | 14.89 | 14.87 | 14.65 | 15.32 | 14.96 |
| 2455180.54511 | 15.31 | 0.04 | MASTER-Vostryakovo | 14.43 | 14.93 | 14.88 | 14.85 | 14.85 | 14.58 | 15.43 | 15.26 |
| 2455180.54599 | 15.34 | 0.09 | MASTER-Vostryakovo | 14.49 | 14.92 | 14.85 | 14.76 | 14.85 | 14.60 | 15.33 | 15.38 |
| 2455181.41010 | 15.31 | 0.07 | MASTER-Vostryakovo | 14.39 | 15.04 | 14.86 | 14.83 | 14.92 | 14.66 | 15.43 | 15.02 |
| 2455181.41559 | 15.42 | 0.06 | MASTER-Vostryakovo | 14.45 | 14.84 | 14.82 | 14.80 | 14.89 | 14.65 | 15.55 | 15.21 |
| 2455181.42486 | 15.43 | 0.06 | MASTER-Vostryakovo | 14.35 | 14.99 | 14.87 | 14.85 | 14.99 | 14.59 | 15.46 | 15.08 |
| 2455181.49634 | 15.46 | 0.05 | MASTER-Vostryakovo | 14.41 | 14.91 | 14.95 | 14.91 | 14.82 | 14.61 | 15.44 | 15.16 |
| 2455181.49785 | 15.43 | 0.10 | MASTER-Vostryakovo | 14.43 | 14.79 | 14.88 | 14.77 | 14.87 | 14.61 | 15.58 | 15.36 |
| 2455184.36704 | 15.75 | 0.07 | MASTER-Vostryakovo | 14.43 | 15.01 | 14.80 | 14.80 | 15.01 | 14.53 | 15.43 | 15.17 |
| 2455184.36803 | 15.60 | 0.07 | MASTER-Vostryakovo | 14.44 | 14.87 | 14.97 | 14.84 | 14.76 | 14.62 | 15.42 | 15.29 |
| 2455184.36905 | 15.68 | 0.05 | MASTER-Vostryakovo | 14.42 | 14.89 | 14.86 | 14.86 | 14.87 | 14.71 | 15.47 | 15.07 |
| 2455208.33413 | 15.29 | 0.06 | MASTER-Vostryakovo | 14.45 | 14.96 | 14.98 | 14.81 | 14.83 | 14.63 | 15.31 | 15.16 |
| 2455208.33589 | 15.26 | 0.04 | MASTER-Vostryakovo | 14.46 | 14.95 | 14.85 | 14.83 | 14.89 | 14.64 | 15.44 | 15.07 |
| 2455208.33764 | 15.25 | 0.08 | MASTER-Vostryakovo | 14.39 | 14.85 | 14.80 | 14.86 | 14.93 | 14.59 | 15.57 | 15.31 |

| | | | | | | | | | | |
|---|---|---|---|---|---|---|---|---|---|---|
| 2455216.27511 | 15.22 | 0.02 | MASTER-Vostryakovo | 14.43 | 14.89 | 14.87 | 14.81 | 14.89 | 14.66 | 15.45 | 15.16 |
| 2455216.27689 | 15.26 | 0.02 | MASTER-Vostryakovo | 14.43 | 14.93 | 14.88 | 14.80 | 14.87 | 14.66 | 15.45 | 15.15 |
| 2455216.27783 | 15.24 | 0.02 | MASTER-Vostryakovo | 14.45 | 14.91 | 14.88 | 14.82 | 14.91 | 14.61 | 15.41 | 15.18 |
| 2455216.34619 | 15.24 | 0.03 | MASTER-Vostryakovo | 14.45 | 14.94 | 14.89 | 14.86 | 14.85 | 14.62 | 15.41 | 15.14 |
| 2455216.34713 | 15.27 | 0.03 | MASTER-Vostryakovo | 14.43 | 14.91 | 14.88 | 14.86 | 14.84 | 14.61 | 15.49 | 15.19 |
| 2455216.34892 | 15.19 | 0.04 | MASTER-Vostryakovo | 14.41 | 14.94 | 14.86 | 14.85 | 14.93 | 14.56 | 15.43 | 15.25 |
| 2455234.29558 | 15.30 | 0.02 | MASTER-Vostryakovo | 14.41 | 14.92 | 14.86 | 14.82 | 14.87 | 14.67 | 15.48 | 15.18 |
| 2455234.29649 | 15.33 | 0.04 | MASTER-Vostryakovo | 14.44 | 14.87 | 14.87 | 14.87 | 14.86 | 14.59 | 15.49 | 15.22 |
| 2455234.29740 | 15.32 | 0.04 | MASTER-Vostryakovo | 14.46 | 14.97 | 14.88 | 14.83 | 14.88 | 14.61 | 15.35 | 15.16 |
| 2455234.34672 | 15.35 | 0.02 | MASTER-Vostryakovo | 14.46 | 14.94 | 14.84 | 14.85 | 14.85 | 14.61 | 15.43 | 15.18 |
| 2455234.34763 | 15.26 | 0.04 | MASTER-Vostryakovo | 14.39 | 14.92 | 14.93 | 14.82 | 14.89 | 14.68 | 15.39 | 15.17 |
| 2455234.34854 | 15.35 | 0.04 | MASTER-Vostryakovo | 14.43 | 14.92 | 14.88 | 14.81 | 14.94 | 14.57 | 15.51 | 15.16 |
| 2455501.49600 | 14.97 | 0.02 | MASTER-Kislovodsk | 14.43 | 14.91 | 14.89 | 14.84 | 14.92 | 14.62 | 15.43 | 15.16 |
| 2455575.22986 | 14.83 | 0.01 | MASTER-Kislovodsk | 14.43 | 14.93 | 14.88 | 14.82 | 14.86 | 14.62 | 15.46 | 15.19 |
| 2455575.27024 | 14.84 | 0.01 | MASTER-Kislovodsk | 14.44 | 14.93 | 14.87 | 14.84 | 14.87 | 14.63 | 15.44 | 15.18 |
| 2455583.02698 | 14.68 | 0.01 | MASTER-Tunka | 14.44 | 14.92 | 14.87 | 14.81 | 14.88 | 14.63 | 15.45 | 15.18 |
| 2456196.54719 | 14.96 | 0.01 | MASTER-Kislovodsk | 14.43 | 14.92 | 14.88 | 14.83 | 14.88 | 14.61 | 15.45 | 15.19 |
| 2456196.54719 | 14.96 | 0.01 | MASTER-Kislovodsk | 14.43 | 14.92 | 14.88 | 14.83 | 14.88 | 14.61 | 15.45 | 15.18 |
| 2456196.57887 | 14.98 | 0.01 | MASTER-Kislovodsk | 14.43 | 14.92 | 14.87 | 14.84 | 14.89 | 14.63 | 15.43 | 15.17 |
| 2456196.57887 | 14.98 | 0.01 | MASTER-Kislovodsk | 14.43 | 14.92 | 14.87 | 14.84 | 14.89 | 14.63 | 15.43 | 15.17 |
| 2456248.45058 | 15.17 | 0.01 | MASTER-Kislovodsk | 14.44 | 14.91 | 14.89 | 14.83 | 14.89 | 14.62 | 15.44 | 15.16 |
| 2456248.46963 | 15.15 | 0.01 | MASTER-Kislovodsk | 14.42 | 14.92 | 14.88 | 14.85 | 14.90 | 14.61 | 15.43 | 15.18 |
| 2456248.48881 | 15.16 | 0.01 | MASTER-Kislovodsk | 14.43 | 14.93 | 14.87 | 14.82 | 14.86 | 14.63 | 15.45 | 15.20 |
| 2456302.28904 | 15.11 | 0.01 | MASTER-Kislovodsk | 14.44 | 14.92 | 14.88 | 14.85 | 14.85 | 14.62 | 15.44 | 15.19 |
| 2456302.29027 | 15.12 | 0.02 | MASTER-Kislovodsk | 14.43 | 14.92 | 14.90 | 14.84 | 14.86 | 14.64 | 15.43 | 15.18 |
| 2456302.32323 | 15.12 | 0.01 | MASTER-Kislovodsk | 14.42 | 14.93 | 14.87 | 14.82 | 14.89 | 14.62 | 15.45 | 15.19 |
| 2456302.32837 | 15.10 | 0.01 | MASTER-Kislovodsk | 14.42 | 14.92 | 14.88 | 14.84 | 14.90 | 14.62 | 15.44 | 15.17 |
| 2456302.32949 | 15.13 | 0.02 | MASTER-Kislovodsk | 14.45 | 14.93 | 14.86 | 14.82 | 14.92 | 14.61 | 15.44 | 15.17 |
| 2456332.21278 | 15.01 | 0.02 | MASTER-Kislovodsk | 14.42 | 14.94 | 14.87 | 14.84 | 14.91 | 14.61 | 15.40 | 15.18 |
| 2456332.21278 | 15.05 | 0.02 | MASTER-Kislovodsk | 14.46 | 14.92 | 14.89 | 14.83 | 14.83 | 14.63 | 15.46 | 15.16 |
| 2456332.24309 | 15.01 | 0.02 | MASTER-Kislovodsk | 14.42 | 14.92 | 14.87 | 14.81 | 14.90 | 14.61 | 15.47 | 15.18 |
| 2456332.24309 | 15.06 | 0.02 | MASTER-Kislovodsk | 14.43 | 14.92 | 14.91 | 14.84 | 14.88 | 14.61 | 15.40 | 15.20 |
| 2456358.18079 | 14.98 | 0.01 | MASTER-Kislovodsk | 14.41 | 14.93 | 14.86 | 14.85 | 14.88 | 14.63 | 15.45 | 15.18 |
| 2456358.21258 | 14.97 | 0.01 | MASTER-Kislovodsk | 14.44 | 14.90 | 14.86 | 14.83 | 14.89 | 14.64 | 15.46 | 15.17 |
| 2456630.17003 | 14.39 | 0.02 | MASTER-Tunka | 14.43 | 14.90 | 14.89 | 14.84 | 14.89 | 14.63 | 15.45 | 15.16 |
| 2456630.17130 | 14.41 | 0.01 | MASTER-Tunka | 14.44 | 14.93 | 14.88 | 14.84 | 14.87 | 14.61 | 15.43 | 15.19 |
| 2456630.20090 | 14.39 | 0.01 | MASTER-Tunka | 14.44 | 14.93 | 14.88 | 14.82 | 14.87 | 14.61 | 15.45 | 15.20 |
| 2456630.20262 | 14.41 | 0.01 | MASTER-Tunka | 14.42 | 14.94 | 14.88 | 14.84 | 14.88 | 14.61 | 15.45 | 15.17 |
| 2456630.20386 | 14.39 | 0.01 | MASTER-Tunka | 14.43 | 14.93 | 14.88 | 14.83 | 14.89 | 14.62 | 15.43 | 15.17 |
| 2456663.07550 | 14.60 | 0.02 | MASTER-Amur | 14.41 | 14.90 | 14.90 | 14.87 | 14.88 | 14.62 | 15.40 | 15.19 |
| 2456663.07673 | 14.62 | 0.03 | MASTER-Amur | 14.40 | 14.90 | 14.91 | 14.85 | 14.88 | 14.66 | 15.43 | 15.15 |
| 2456663.07795 | 14.64 | 0.02 | MASTER-Amur | 14.40 | 14.93 | 14.87 | 14.84 | 14.90 | 14.63 | 15.43 | 15.20 |
| 2456936.23566 | 14.86 | 0.04 | MASTER-Amur | 14.44 | 14.92 | 14.88 | 14.80 | 14.86 | 14.61 | 15.55 | 15.17 |
| 2456936.23566 | 14.83 | 0.01 | MASTER-Amur | 14.45 | 14.92 | 14.86 | 14.83 | 14.87 | 14.61 | 15.46 | 15.19 |
| 2456936.23666 | 14.88 | 0.02 | MASTER-Amur | 14.44 | 14.91 | 14.83 | 14.85 | 14.88 | 14.62 | 15.45 | 15.22 |
| 2456936.23666 | 14.82 | 0.01 | MASTER-Amur | 14.43 | 14.92 | 14.88 | 14.83 | 14.89 | 14.64 | 15.43 | 15.17 |
| 2456936.23766 | 14.87 | 0.01 | MASTER-Amur | 14.44 | 14.93 | 14.89 | 14.81 | 14.88 | 14.60 | 15.44 | 15.18 |
| 2456936.23766 | 14.83 | 0.01 | MASTER-Amur | 14.43 | 14.92 | 14.89 | 14.84 | 14.89 | 14.61 | 15.43 | 15.17 |
| 2456936.26682 | 14.89 | 0.02 | MASTER-Amur | 14.47 | 14.92 | 14.87 | 14.84 | 14.90 | 14.58 | 15.44 | 15.15 |
| 2456936.26782 | 14.83 | 0.02 | MASTER-Amur | 14.43 | 14.95 | 14.89 | 14.82 | 14.91 | 14.61 | 15.39 | 15.18 |
| 2456936.26883 | 14.88 | 0.02 | MASTER-Amur | 14.46 | 14.94 | 14.87 | 14.81 | 14.85 | 14.64 | 15.42 | 15.16 |
| 2456974.37371 | 15.00 | 0.02 | MASTER-Tunka | 14.43 | 14.93 | 14.88 | 14.83 | 14.92 | 14.63 | 15.41 | 15.14 |
| 2456974.37494 | 14.99 | 0.02 | MASTER-Tunka | 14.43 | 14.95 | 14.85 | 14.83 | 14.87 | 14.66 | 15.40 | 15.18 |
| 2456974.37615 | 15.02 | 0.03 | MASTER-Tunka | 14.43 | 14.95 | 14.86 | 14.83 | 14.87 | 14.64 | 15.39 | 15.21 |
| 2456974.37742 | 15.03 | 0.01 | MASTER-Tunka | 14.43 | 14.91 | 14.87 | 14.83 | 14.88 | 14.63 | 15.44 | 15.20 |

| | | | | | | | | | | |
|---|---|---|---|---|---|---|---|---|---|---|
| 2456974.37859 | 14.98 | 0.02 | MASTER-Tunka | 14.42 | 14.95 | 14.84 | 14.85 | 14.91 | 14.63 | 15.43 | 15.15 |
| 2456974.38095 | 15.04 | 0.02 | MASTER-Tunka | 14.44 | 14.93 | 14.85 | 14.85 | 14.90 | 14.59 | 15.47 | 15.16 |
| 2456974.38213 | 15.00 | 0.02 | MASTER-Tunka | 14.44 | 14.95 | 14.86 | 14.82 | 14.91 | 14.62 | 15.44 | 15.15 |
| 2456974.38332 | 15.00 | 0.02 | MASTER-Tunka | 14.44 | 14.91 | 14.86 | 14.87 | 14.87 | 14.61 | 15.42 | 15.20 |
| 2456974.38451 | 15.05 | 0.02 | MASTER-Tunka | 14.43 | 14.91 | 14.86 | 14.86 | 14.91 | 14.60 | 15.41 | 15.19 |
| 2456974.38571 | 14.99 | 0.03 | MASTER-Tunka | 14.44 | 14.95 | 14.88 | 14.77 | 14.86 | 14.63 | 15.50 | 15.17 |
| 2456974.45690 | 15.00 | 0.06 | MASTER-Tunka | 14.43 | 14.97 | 14.94 | 14.77 | 14.88 | 14.60 | 15.33 | 15.25 |
| 2456974.45813 | 14.92 | 0.05 | MASTER-Tunka | 14.40 | 14.88 | 14.90 | 14.84 | 14.90 | 14.65 | 15.54 | 15.11 |
| 2456974.45932 | 15.07 | 0.03 | MASTER-Tunka | 14.46 | 14.89 | 14.87 | 14.88 | 14.81 | 14.63 | 15.45 | 15.17 |
| 2456974.46050 | 15.12 | 0.06 | MASTER-Tunka | 14.43 | 14.84 | 14.96 | 14.83 | 14.87 | 14.58 | 15.54 | 15.22 |
| 2457128.21489 | 14.94 | 0.08 | MASTER-Kislovodsk | 14.37 | 14.98 | 15.00 | 14.81 | 14.75 | 14.49 | 15.42 | 15.17 |
| 2457128.22050 | 14.79 | 0.04 | MASTER-Kislovodsk | 14.50 | 14.91 | 14.84 | 14.78 | 14.76 | 14.61 | 15.51 | 15.27 |
| 2457128.22881 | 14.81 | 0.06 | MASTER-Kislovodsk | 14.46 | 14.87 | 15.02 | 14.92 | 14.75 | 14.75 | 15.45 | 15.17 |
| 2457128.22982 | 14.90 | 0.04 | MASTER-Kislovodsk | 14.39 | 15.03 | 14.86 | 14.84 | 14.80 | 14.63 | 15.42 | 15.23 |
| 2457128.23087 | 14.95 | 0.04 | MASTER-Kislovodsk | 14.50 | 14.86 | 14.82 | 14.92 | 14.78 | 14.67 | 15.53 | 15.10 |
| 2457128.23193 | 14.83 | 0.05 | MASTER-Kislovodsk | 14.53 | 14.92 | 14.81 | 14.90 | 14.69 | 14.70 | 15.36 | 15.28 |
| 2457229.66344 | 15.70 | 0.01 | MASTER-SAAO | 14.44 | 14.93 | 14.89 | 14.82 | 14.87 | 14.64 | 15.43 | 15.18 |
| 2457229.66792 | 15.69 | 0.01 | MASTER-SAAO | 14.44 | 14.92 | 14.87 | 14.84 | 14.88 | 14.61 | 15.44 | 15.18 |
| 2457229.67369 | 15.70 | 0.01 | MASTER-SAAO | 14.42 | 14.92 | 14.88 | 14.84 | 14.90 | 14.61 | 15.45 | 15.18 |
| 2457304.48348 | 15.51 | 0.01 | MASTER-Kislovodsk | 14.44 | 14.91 | 14.86 | 14.85 | 14.87 | 14.63 | 15.45 | 15.16 |
| 2457304.49856 | 15.49 | 0.01 | MASTER-Kislovodsk | 14.43 | 14.92 | 14.88 | 14.83 | 14.88 | 14.64 | 15.45 | 15.17 |
| 2457374.27641 | 15.34 | 0.02 | MASTER-Kislovodsk | 14.43 | 14.93 | 14.89 | 14.81 | 14.88 | 14.60 | 15.44 | 15.21 |
| 2457374.28383 | 15.35 | 0.01 | MASTER-Kislovodsk | 14.43 | 14.92 | 14.88 | 14.85 | 14.89 | 14.61 | 15.43 | 15.18 |
| 2457494.21180 | 15.23 | 0.02 | MASTER-Kislovodsk | 14.42 | 14.93 | 14.90 | 14.84 | 14.90 | 14.62 | 15.42 | 15.16 |
| 2457494.22071 | 15.24 | 0.02 | MASTER-Kislovodsk | 14.41 | 14.93 | 14.88 | 14.80 | 14.90 | 14.60 | 15.44 | 15.23 |
| 2457639.60975 | 14.65 | 0.01 | MASTER-SAAO | 14.43 | 14.92 | 14.88 | 14.83 | 14.88 | 14.62 | 15.44 | 15.18 |
| 2457639.61773 | 14.65 | 0.01 | MASTER-SAAO | 14.43 | 14.92 | 14.87 | 14.83 | 14.89 | 14.62 | 15.45 | 15.17 |
| 2457687.82814 | 14.83 | 0.01 | MASTER-OAFA | 14.43 | 14.92 | 14.88 | 14.84 | 14.88 | 14.61 | 15.45 | 15.17 |
| 2457687.84631 | 14.85 | 0.01 | MASTER-OAFA | 14.43 | 14.92 | 14.87 | 14.83 | 14.89 | 14.62 | 15.43 | 15.19 |
| 2457687.85665 | 14.83 | 0.01 | MASTER-OAFA | 14.43 | 14.93 | 14.88 | 14.82 | 14.88 | 14.63 | 15.44 | 15.18 |
| 2457729.30139 | 14.99 | 0.01 | MASTER-Kislovodsk | 14.43 | 14.93 | 14.88 | 14.83 | 14.88 | 14.61 | 15.44 | 15.18 |
| 2457729.32019 | 14.96 | 0.01 | MASTER-Kislovodsk | 14.43 | 14.92 | 14.86 | 14.84 | 14.88 | 14.62 | 15.46 | 15.16 |
| 2457752.42302 | 15.19 | 0.01 | MASTER-SAAO | 14.43 | 14.91 | 14.87 | 14.85 | 14.89 | 14.62 | 15.44 | 15.18 |
| 2457752.42664 | 15.19 | 0.01 | MASTER-SAAO | 14.44 | 14.93 | 14.88 | 14.82 | 14.88 | 14.62 | 15.44 | 15.18 |
| 2457822.52884 | 14.89 | 0.02 | MASTER-OAFA | 14.43 | 14.90 | 14.87 | 14.86 | 14.92 | 14.59 | 15.47 | 15.18 |
| 2457822.53934 | 14.85 | 0.02 | MASTER-OAFA | 14.41 | 14.92 | 14.89 | 14.83 | 14.90 | 14.62 | 15.48 | 15.16 |
| 2457822.54724 | 14.90 | 0.02 | MASTER-OAFA | 14.43 | 14.91 | 14.91 | 14.82 | 14.90 | 14.60 | 15.45 | 15.18 |
| 2457827.60007 | 15.00 | 0.02 | MASTER-OAFA | 14.42 | 14.94 | 14.86 | 14.84 | 14.86 | 14.65 | 15.42 | 15.18 |
| 2457878.49282 | 14.85 | 0.03 | MASTER-OAFA | 14.46 | 14.95 | 14.87 | 14.81 | 14.84 | 14.65 | 15.38 | 15.20 |
| 2458013.43685 | 14.12 | 0.02 | MASTER-Kislovodsk | 14.44 | 14.92 | 14.88 | 14.85 | 14.87 | 14.64 | 15.42 | 15.16 |
| 2458013.44114 | 14.10 | 0.01 | MASTER-Kislovodsk | 14.45 | 14.91 | 14.87 | 14.82 | 14.86 | 14.63 | 15.46 | 15.19 |
| 2458019.37307 | 15.13 | 0.01 | MASTER-Tavrida | 14.44 | 14.92 | 14.87 | 14.83 | 14.90 | 14.61 | 15.44 | 15.17 |
| 2458019.37544 | 15.12 | 0.01 | MASTER-Tavrida | 14.43 | 14.92 | 14.88 | 14.83 | 14.87 | 14.62 | 15.44 | 15.20 |
| 2458019.37779 | 15.11 | 0.01 | MASTER-Tavrida | 14.43 | 14.92 | 14.88 | 14.84 | 14.88 | 14.63 | 15.45 | 15.17 |
| 2458019.46743 | 14.35 | 0.01 | MASTER-Tavrida | 14.44 | 14.92 | 14.89 | 14.81 | 14.88 | 14.62 | 15.44 | 15.19 |
| 2458019.47598 | 14.31 | 0.02 | MASTER-Tavrida | 14.42 | 14.95 | 14.90 | 14.84 | 14.87 | 14.59 | 15.45 | 15.18 |
| 2458019.47853 | 14.33 | 0.01 | MASTER-Tavrida | 14.43 | 14.94 | 14.86 | 14.82 | 14.88 | 14.63 | 15.43 | 15.20 |
| 2458019.48381 | 14.33 | 0.01 | MASTER-Tavrida | 14.45 | 14.93 | 14.90 | 14.82 | 14.86 | 14.61 | 15.44 | 15.17 |
| 2458019.48635 | 14.33 | 0.01 | MASTER-Tavrida | 14.44 | 14.93 | 14.88 | 14.82 | 14.88 | 14.61 | 15.44 | 15.18 |
| 2458019.48887 | 14.29 | 0.03 | MASTER-Tavrida | 14.41 | 14.85 | 14.90 | 14.84 | 14.90 | 14.67 | 15.45 | 15.17 |
| 2458019.49410 | 14.31 | 0.01 | MASTER-Tavrida | 14.42 | 14.91 | 14.89 | 14.85 | 14.89 | 14.62 | 15.44 | 15.18 |
| 2458019.49943 | 14.34 | 0.03 | MASTER-Tavrida | 14.45 | 14.95 | 14.82 | 14.86 | 14.90 | 14.61 | 15.44 | 15.16 |
| 2458019.50484 | 14.34 | 0.01 | MASTER-Tavrida | 14.43 | 14.92 | 14.88 | 14.82 | 14.90 | 14.63 | 15.43 | 15.17 |
| 2458020.49157 | 14.40 | 0.01 | MASTER-Tavrida | 14.42 | 14.93 | 14.87 | 14.84 | 14.89 | 14.61 | 15.45 | 15.19 |
| 2458020.49414 | 14.41 | 0.01 | MASTER-Tavrida | 14.44 | 14.92 | 14.88 | 14.84 | 14.86 | 14.62 | 15.44 | 15.18 |
| 2458020.50408 | 14.41 | 0.01 | MASTER-Tavrida | 14.43 | 14.93 | 14.88 | 14.82 | 14.88 | 14.61 | 15.43 | 15.19 |
| 2458020.50714 | 14.41 | 0.02 | MASTER-Tavrida | 14.47 | 14.91 | 14.89 | 14.84 | 14.86 | 14.61 | 15.42 | 15.16 |
| 2458020.52273 | 14.40 | 0.02 | MASTER-Tavrida | 14.43 | 14.96 | 14.88 | 14.84 | 14.87 | 14.62 | 15.43 | 15.16 |
| 2458020.52525 | 14.40 | 0.01 | MASTER-Tavrida | 14.44 | 14.91 | 14.85 | 14.83 | 14.90 | 14.62 | 15.46 | 15.18 |
| 2458020.52780 | 14.40 | 0.02 | MASTER-Tavrida | 14.41 | 14.91 | 14.89 | 14.83 | 14.89 | 14.63 | 15.47 | 15.18 |
| 2458020.54335 | 14.41 | 0.01 | MASTER-Tavrida | 14.42 | 14.92 | 14.87 | 14.83 | 14.89 | 14.63 | 15.45 | 15.19 |
| 2458020.60758 | 14.39 | 0.01 | MASTER-SAAO | 14.42 | 14.94 | 14.89 | 14.83 | 14.88 | 14.62 | 15.45 | 15.18 |
| 2458020.60758 | 14.40 | 0.01 | MASTER-SAAO | 14.43 | 14.91 | 14.88 | 14.84 | 14.89 | 14.62 | 15.45 | 15.18 |
| 2458020.61011 | 14.40 | 0.01 | MASTER-SAAO | 14.42 | 14.92 | 14.87 | 14.82 | 14.89 | 14.63 | 15.44 | 15.19 |
| 2458020.61011 | 14.41 | 0.01 | MASTER-SAAO | 14.43 | 14.92 | 14.87 | 14.83 | 14.89 | 14.64 | 15.43 | 15.17 |
| 2458020.63066 | 14.42 | 0.01 | MASTER-SAAO | 14.44 | 14.91 | 14.87 | 14.83 | 14.89 | 14.61 | 15.45 | 15.19 |
| 2458020.63066 | 14.41 | 0.01 | MASTER-SAAO | 14.43 | 14.93 | 14.88 | 14.83 | 14.88 | 14.62 | 15.43 | 15.18 |
| 2458020.63350 | 14.42 | 0.01 | MASTER-SAAO | 14.42 | 14.92 | 14.89 | 14.84 | 14.87 | 14.64 | 15.44 | 15.17 |
| 2458020.63350 | 14.41 | 0.01 | MASTER-SAAO | 14.44 | 14.92 | 14.87 | 14.82 | 14.88 | 14.62 | 15.45 | 15.19 |
| 2458020.63914 | 14.42 | 0.01 | MASTER-SAAO | 14.45 | 14.92 | 14.85 | 14.84 | 14.88 | 14.60 | 15.44 | 15.19 |
| 2458020.63914 | 14.41 | 0.01 | MASTER-SAAO | 14.45 | 14.93 | 14.86 | 14.83 | 14.87 | 14.62 | 15.44 | 15.18 |
| 2458020.64289 | 14.41 | 0.01 | MASTER-SAAO | 14.44 | 14.93 | 14.88 | 14.82 | 14.89 | 14.63 | 15.43 | 15.18 |
| 2458020.64289 | 14.41 | 0.01 | MASTER-SAAO | 14.43 | 14.93 | 14.89 | 14.83 | 14.88 | 14.61 | 15.43 | 15.18 |
| 2458020.64724 | 14.40 | 0.01 | MASTER-SAAO | 14.44 | 14.92 | 14.89 | 14.85 | 14.88 | 14.61 | 15.43 | 15.15 |
| 2458020.64724 | 14.42 | 0.01 | MASTER-SAAO | 14.43 | 14.91 | 14.89 | 14.83 | 14.88 | 14.62 | 15.45 | 15.19 |
| 2458026.44319 | 14.52 | 0.01 | MASTER-Tavrida | 14.44 | 14.93 | 14.87 | 14.84 | 14.88 | 14.61 | 15.42 | 15.19 |
| 2458026.44700 | 14.52 | 0.02 | MASTER-Tavrida | 14.44 | 14.94 | 14.89 | 14.81 | 14.88 | 14.60 | 15.42 | 15.19 |
| 2458026.45210 | 14.50 | 0.01 | MASTER-Tavrida | 14.44 | 14.92 | 14.89 | 14.82 | 14.86 | 14.62 | 15.44 | 15.19 |
| 2458026.45456 | 14.49 | 0.01 | MASTER-Tavrida | 14.43 | 14.95 | 14.88 | 14.83 | 14.86 | 14.62 | 15.43 | 15.19 |
| 2458026.45965 | 14.50 | 0.01 | MASTER-Tavrida | 14.43 | 14.93 | 14.88 | 14.83 | 14.85 | 14.63 | 15.44 | 15.20 |
| 2458026.46213 | 14.45 | 0.04 | MASTER-Tavrida | 14.41 | 14.87 | 14.91 | 14.91 | 14.92 | 14.61 | 15.39 | 15.15 |
| 2458026.46463 | 14.51 | 0.01 | MASTER-Tavrida | 14.44 | 14.94 | 14.88 | 14.82 | 14.88 | 14.60 | 15.44 | 15.20 |
| 2458026.46974 | 14.52 | 0.02 | MASTER-Tavrida | 14.46 | 14.91 | 14.88 | 14.82 | 14.86 | 14.61 | 15.45 | 15.20 |
| 2458026.47222 | 14.49 | 0.02 | MASTER-Tavrida | 14.43 | 14.90 | 14.86 | 14.86 | 14.89 | 14.65 | 15.45 | 15.15 |
| 2458026.47469 | 14.53 | 0.01 | MASTER-Tavrida | 14.43 | 14.93 | 14.89 | 14.81 | 14.89 | 14.61 | 15.46 | 15.17 |
| 2458026.47980 | 14.51 | 0.01 | MASTER-Tavrida | 14.43 | 14.92 | 14.86 | 14.82 | 14.89 | 14.63 | 15.46 | 15.17 |
| 2458026.48229 | 14.49 | 0.02 | MASTER-Tavrida | 14.41 | 14.91 | 14.88 | 14.83 | 14.91 | 14.64 | 15.47 | 15.17 |
| 2458026.48477 | 14.49 | 0.02 | MASTER-Tavrida | 14.43 | 14.93 | 14.85 | 14.81 | 14.90 | 14.63 | 15.47 | 15.17 |
| 2458027.77367 | 14.39 | 0.01 | MASTER-OAFA | 14.43 | 14.92 | 14.87 | 14.83 | 14.88 | 14.62 | 15.45 | 15.19 |
| 2458027.77922 | 14.40 | 0.01 | MASTER-OAFA | 14.43 | 14.92 | 14.88 | 14.83 | 14.88 | 14.61 | 15.46 | 15.20 |
| 2458027.78466 | 14.37 | 0.01 | MASTER-OAFA | 14.42 | 14.94 | 14.88 | 14.82 | 14.90 | 14.61 | 15.46 | 15.17 |
| 2458027.79003 | 14.38 | 0.01 | MASTER-OAFA | 14.45 | 14.92 | 14.88 | 14.83 | 14.86 | 14.63 | 15.45 | 15.16 |
| 2458027.79535 | 14.40 | 0.02 | MASTER-OAFA | 14.44 | 14.92 | 14.88 | 14.84 | 14.90 | 14.62 | 15.40 | 15.17 |
| 2458027.80281 | 14.40 | 0.01 | MASTER-OAFA | 14.42 | 14.92 | 14.88 | 14.84 | 14.88 | 14.64 | 15.43 | 15.19 |
| 2458040.39755 | 14.45 | 0.01 | MASTER-Kislovodsk | 14.44 | 14.94 | 14.87 | 14.83 | 14.89 | 14.60 | 15.44 | 15.18 |
| 2458040.39756 | 14.43 | 0.01 | MASTER-Kislovodsk | 14.42 | 14.94 | 14.87 | 14.85 | 14.87 | 14.63 | 15.42 | 15.19 |
| 2458040.40331 | 14.44 | 0.03 | MASTER-Kislovodsk | 14.44 | 14.86 | 14.83 | 14.84 | 14.88 | 14.63 | 15.48 | 15.22 |
| 2458040.40331 | 14.45 | 0.02 | MASTER-Kislovodsk | 14.44 | 14.90 | 14.90 | 14.83 | 14.85 | 14.63 | 15.44 | 15.18 |
| 2458040.41137 | 14.47 | 0.02 | MASTER-Kislovodsk | 14.43 | 14.90 | 14.87 | 14.83 | 14.89 | 14.61 | 15.49 | 15.17 |
| 2458040.41137 | 14.47 | 0.03 | MASTER-Kislovodsk | 14.44 | 14.96 | 14.88 | 14.83 | 14.90 | 14.64 | 15.39 | 15.15 |
| 2458040.41720 | 14.45 | 0.01 | MASTER-Kislovodsk | 14.41 | 14.94 | 14.89 | 14.85 | 14.88 | 14.62 | 15.43 | 15.17 |
| 2458040.41720 | 14.45 | 0.02 | MASTER-Kislovodsk | 14.44 | 14.95 | 14.90 | 14.83 | 14.88 | 14.59 | 15.44 | 15.15 |
| 2458040.42288 | 14.45 | 0.02 | MASTER-Kislovodsk | 14.43 | 14.92 | 14.89 | 14.82 | 14.89 | 14.58 | 15.46 | 15.20 |
| 2458040.42288 | 14.46 | 0.02 | MASTER-Kislovodsk | 14.42 | 14.91 | 14.86 | 14.81 | 14.86 | 14.65 | 15.48 | 15.19 |
| 2458040.42850 | 14.43 | 0.03 | MASTER-Kislovodsk | 14.44 | 14.93 | 14.88 | 14.82 | 14.91 | 14.62 | 15.38 | 15.20 |
| 2458040.42850 | 14.46 | 0.02 | MASTER-Kislovodsk | 14.43 | 14.92 | 14.88 | 14.85 | 14.87 | 14.64 | 15.44 | 15.14 |
| 2458042.19633 | 14.47 | 0.02 | MASTER-Amur | 14.45 | 14.91 | 14.89 | 14.84 | 14.85 | 14.64 | 15.40 | 15.16 |
| 2458042.20117 | 14.41 | 0.03 | MASTER-Amur | 14.41 | 14.94 | 14.87 | 14.87 | 14.91 | 14.60 | 15.37 | 15.21 |
| 2458042.20362 | 14.41 | 0.04 | MASTER-Amur | 14.38 | 14.91 | 14.93 | 14.84 | 14.92 | 14.65 | 15.47 | 15.10 |
| 2458042.20604 | 14.46 | 0.04 | MASTER-Amur | 14.43 | 14.93 | 14.82 | 14.79 | 14.89 | 14.66 | 15.42 | 15.24 |
| 2458042.21391 | 14.42 | 0.03 | MASTER-Amur | 14.45 | 14.97 | 14.89 | 14.85 | 14.84 | 14.58 | 15.47 | 15.15 |
| 2458042.21878 | 14.43 | 0.02 | MASTER-Amur | 14.42 | 14.90 | 14.89 | 14.81 | 14.88 | 14.62 | 15.47 | 15.21 |
| 2458042.22363 | 14.44 | 0.03 | MASTER-Amur | 14.48 | 14.90 | 14.85 | 14.81 | 14.88 | 14.60 | 15.49 | 15.18 |
| 2458043.24063 | 14.32 | 0.03 | MASTER-Amur | 14.42 | 14.93 | 14.89 | 14.86 | 14.88 | 14.65 | 15.45 | 15.11 |
| 2458043.24550 | 14.34 | 0.03 | MASTER-Amur | 14.42 | 14.90 | 14.88 | 14.80 | 14.93 | 14.59 | 15.50 | 15.21 |
| 2458043.24795 | 14.37 | 0.04 | MASTER-Amur | 14.44 | 14.96 | 14.87 | 14.80 | 14.84 | 14.63 | 15.40 | 15.25 |
| 2458043.25038 | 14.32 | 0.04 | MASTER-Amur | 14.42 | 14.93 | 14.89 | 14.85 | 14.94 | 14.62 | 15.45 | 15.08 |
| 2458043.26063 | 14.41 | 0.04 | MASTER-Amur | 14.46 | 14.90 | 14.87 | 14.85 | 14.82 | 14.61 | 15.41 | 15.25 |
| 2458043.38586 | 14.37 | 0.02 | MASTER-Tavrida | 14.41 | 14.93 | 14.88 | 14.82 | 14.91 | 14.63 | 15.42 | 15.19 |
| 2458043.42512 | 14.36 | 0.01 | MASTER-Tavrida | 14.44 | 14.91 | 14.87 | 14.83 | 14.89 | 14.63 | 15.44 | 15.17 |
| 2458043.42759 | 14.39 | 0.01 | MASTER-Tavrida | 14.45 | 14.91 | 14.86 | 14.83 | 14.88 | 14.61 | 15.45 | 15.19 |
| 2458043.43014 | 14.36 | 0.01 | MASTER-Tavrida | 14.42 | 14.93 | 14.88 | 14.82 | 14.88 | 14.65 | 15.45 | 15.16 |
| 2458043.44675 | 14.35 | 0.01 | MASTER-Tavrida | 14.43 | 14.91 | 14.87 | 14.83 | 14.89 | 14.63 | 15.46 | 15.17 |
| 2458043.44675 | 14.35 | 0.01 | MASTER-Tavrida | 14.43 | 14.93 | 14.90 | 14.83 | 14.88 | 14.61 | 15.44 | 15.18 |
| 2458043.45696 | 14.35 | 0.01 | MASTER-Tavrida | 14.43 | 14.93 | 14.89 | 14.83 | 14.88 | 14.61 | 15.42 | 15.18 |
| 2458043.45696 | 14.37 | 0.02 | MASTER-Tavrida | 14.44 | 14.92 | 14.87 | 14.86 | 14.85 | 14.61 | 15.44 | 15.19 |
| 2458043.51896 | 14.38 | 0.03 | MASTER-Tavrida | 14.45 | 14.91 | 14.88 | 14.79 | 14.88 | 14.66 | 15.40 | 15.19 |
| 2458044.39100 | 14.36 | 0.02 | MASTER-Tavrida | 14.42 | 14.94 | 14.88 | 14.84 | 14.86 | 14.64 | 15.46 | 15.16 |
| 2458044.39948 | 14.37 | 0.02 | MASTER-Tavrida | 14.43 | 14.90 | 14.86 | 14.87 | 14.90 | 14.64 | 15.42 | 15.15 |
| 2458044.40195 | 14.35 | 0.01 | MASTER-Tavrida | 14.41 | 14.92 | 14.87 | 14.83 | 14.90 | 14.63 | 15.44 | 15.20 |
| 2458044.41267 | 14.35 | 0.01 | MASTER-Tavrida | 14.44 | 14.93 | 14.88 | 14.82 | 14.88 | 14.60 | 15.44 | 15.21 |
| 2458044.41513 | 14.37 | 0.02 | MASTER-Tavrida | 14.44 | 14.94 | 14.89 | 14.83 | 14.86 | 14.59 | 15.45 | 15.20 |
| 2458044.43886 | 14.36 | 0.02 | MASTER-Tavrida | 14.44 | 14.91 | 14.87 | 14.86 | 14.87 | 14.63 | 15.44 | 15.16 |
| 2458044.44136 | 14.34 | 0.01 | MASTER-Tavrida | 14.43 | 14.93 | 14.89 | 14.82 | 14.89 | 14.61 | 15.43 | 15.19 |
| 2458044.44707 | 14.37 | 0.01 | MASTER-Tavrida | 14.43 | 14.92 | 14.87 | 14.82 | 14.92 | 14.62 | 15.45 | 15.17 |
| 2458044.44963 | 14.37 | 0.01 | MASTER-Tavrida | 14.42 | 14.91 | 14.89 | 14.83 | 14.88 | 14.63 | 15.45 | 15.18 |
| 2458044.45213 | 14.37 | 0.01 | MASTER-Tavrida | 14.45 | 14.90 | 14.87 | 14.85 | 14.87 | 14.62 | 15.45 | 15.18 |
| 2458044.47508 | 14.39 | 0.02 | MASTER-Tavrida | 14.44 | 14.96 | 14.90 | 14.81 | 14.88 | 14.59 | 15.46 | 15.17 |
| 2458044.75623 | 14.37 | 0.01 | MASTER-OAFA | 14.44 | 14.94 | 14.86 | 14.83 | 14.88 | 14.62 | 15.44 | 15.18 |
| 2458044.76156 | 14.38 | 0.01 | MASTER-OAFA | 14.43 | 14.93 | 14.89 | 14.84 | 14.89 | 14.61 | 15.42 | 15.17 |
| 2458044.76703 | 14.38 | 0.01 | MASTER-OAFA | 14.43 | 14.93 | 14.89 | 14.83 | 14.86 | 14.63 | 15.45 | 15.18 |
| 2458044.77236 | 14.38 | 0.01 | MASTER-OAFA | 14.44 | 14.91 | 14.89 | 14.83 | 14.89 | 14.61 | 15.43 | 15.18 |
| 2458044.79414 | 14.39 | 0.01 | MASTER-OAFA | 14.42 | 14.92 | 14.87 | 14.82 | 14.89 | 14.63 | 15.45 | 15.19 |
| 2458044.79964 | 14.39 | 0.01 | MASTER-OAFA | 14.44 | 14.91 | 14.88 | 14.84 | 14.89 | 14.61 | 15.45 | 15.17 |
| 2458047.80139 | 14.59 | 0.01 | MASTER-OAFA | 14.43 | 14.92 | 14.87 | 14.85 | 14.88 | 14.61 | 15.45 | 15.18 |
| 2458047.80669 | 14.61 | 0.01 | MASTER-OAFA | 14.44 | 14.92 | 14.89 | 14.82 | 14.89 | 14.61 | 15.45 | 15.18 |
| 2458047.84873 | 14.60 | 0.01 | MASTER-OAFA | 14.43 | 14.91 | 14.88 | 14.84 | 14.87 | 14.63 | 15.42 | 15.19 |
| 2458047.85130 | 14.60 | 0.01 | MASTER-OAFA | 14.43 | 14.92 | 14.88 | 14.82 | 14.89 | 14.63 | 15.43 | 15.18 |
| 2458047.85388 | 14.61 | 0.01 | MASTER-OAFA | 14.43 | 14.94 | 14.87 | 14.84 | 14.88 | 14.62 | 15.45 | 15.17 |
| 2458225.22364 | 14.47 | 0.01 | MASTER-Kislovodsk | 14.42 | 14.91 | 14.87 | 14.83 | 14.91 | 14.63 | 15.43 | 15.18 |
| 2458225.23653 | 14.47 | 0.02 | MASTER-Kislovodsk | 14.45 | 14.91 | 14.88 | 14.84 | 14.88 | 14.64 | 15.43 | 15.16 |
| 2458384.59435 | 14.61 | 0.03 | MASTER-IAC | 14.43 | 14.95 | 14.87 | 14.87 | 14.86 | 14.64 | 15.44 | 15.12 |
| 2458384.59691 | 14.62 | 0.03 | MASTER-IAC | 14.47 | 14.90 | 14.89 | 14.83 | 14.89 | 14.58 | 15.47 | 15.15 |
| 2458384.62700 | 14.60 | 0.03 | MASTER-IAC | 14.41 | 14.94 | 14.85 | 14.79 | 14.92 | 14.66 | 15.41 | 15.20 |
| 2458384.62814 | 14.60 | 0.02 | MASTER-IAC | 14.41 | 14.89 | 14.88 | 14.87 | 14.89 | 14.61 | 15.47 | 15.19 |
| 2458384.62944 | 14.63 | 0.02 | MASTER-IAC | 14.43 | 14.92 | 14.87 | 14.87 | 14.85 | 14.62 | 15.43 | 15.19 |
| 2458449.21264 | 14.70 | 0.03 | MASTER-Tunka | 14.41 | 14.91 | 14.89 | 14.82 | 14.96 | 14.60 | 15.43 | 15.18 |
| 2458449.28012 | 14.67 | 0.04 | MASTER-Tunka | 14.40 | 14.91 | 14.88 | 14.88 | 14.94 | 14.63 | 15.36 | 15.17 |
| 2458452.66756 | 14.66 | 0.01 | MASTER-IAC | 14.44 | 14.92 | 14.87 | 14.83 | 14.92 | 14.61 | 15.43 | 15.17 |
| 2458452.68065 | 14.63 | 0.02 | MASTER-IAC | 14.43 | 14.93 | 14.87 | 14.83 | 14.92 | 14.60 | 15.44 | 15.17 |
| 2458515.43054 | 14.89 | 0.01 | MASTER-IAC | 14.43 | 14.92 | 14.89 | 14.83 | 14.90 | 14.62 | 15.44 | 15.16 |
| 2458515.43428 | 14.88 | 0.02 | MASTER-IAC | 14.42 | 14.92 | 14.87 | 14.83 | 14.93 | 14.61 | 15.44 | 15.17 |
| 2458568.27681 | 14.53 | 0.02 | MASTER-Kislovodsk | 14.43 | 14.92 | 14.87 | 14.86 | 14.87 | 14.60 | 15.47 | 15.17 |
| 2458568.27935 | 14.51 | 0.02 | MASTER-Kislovodsk | 14.41 | 14.93 | 14.89 | 14.82 | 14.89 | 14.63 | 15.41 | 15.21 |
| 2458568.28440 | 14.54 | 0.02 | MASTER-Kislovodsk | 14.44 | 14.94 | 14.86 | 14.80 | 14.88 | 14.63 | 15.45 | 15.19 |
| 2458568.28692 | 14.51 | 0.02 | MASTER-Kislovodsk | 14.44 | 14.92 | 14.87 | 14.81 | 14.91 | 14.60 | 15.48 | 15.15 |
| 2458568.28933 | 14.51 | 0.03 | MASTER-Kislovodsk | 14.43 | 14.93 | 14.90 | 14.86 | 14.83 | 14.61 | 15.41 | 15.20 |
| 2458702.86663 | 14.60 | 0.01 | MASTER-OAFA | 14.43 | 14.92 | 14.88 | 14.82 | 14.88 | 14.62 | 15.46 | 15.21 |
| 2458702.86973 | 14.59 | 0.02 | MASTER-OAFA | 14.43 | 14.92 | 14.86 | 14.83 | 14.89 | 14.61 | 15.49 | 15.19 |
| 2458702.87227 | 14.57 | 0.02 | MASTER-OAFA | 14.42 | 14.91 | 14.89 | 14.84 | 14.86 | 14.65 | 15.43 | 15.19 |
| 2458702.87484 | 14.58 | 0.02 | MASTER-OAFA | 14.45 | 14.93 | 14.85 | 14.83 | 14.87 | 14.60 | 15.44 | 15.21 |
| 2458702.87741 | 14.60 | 0.01 | MASTER-OAFA | 14.43 | 14.92 | 14.89 | 14.84 | 14.88 | 14.62 | 15.42 | 15.17 |
| 2458702.87996 | 14.60 | 0.02 | MASTER-OAFA | 14.44 | 14.92 | 14.84 | 14.85 | 14.88 | 14.63 | 15.43 | 15.17 |
| 2458702.88254 | 14.60 | 0.01 | MASTER-OAFA | 14.43 | 14.95 | 14.86 | 14.82 | 14.88 | 14.64 | 15.43 | 15.19 |
| 2458702.88511 | 14.58 | 0.01 | MASTER-OAFA | 14.43 | 14.92 | 14.89 | 14.82 | 14.88 | 14.62 | 15.44 | 15.19 |
| 2458702.88773 | 14.57 | 0.01 | MASTER-OAFA | 14.43 | 14.92 | 14.87 | 14.83 | 14.88 | 14.62 | 15.45 | 15.19 |
| 2458702.89029 | 14.56 | 0.02 | MASTER-OAFA | 14.41 | 14.91 | 14.93 | 14.84 | 14.88 | 14.61 | 15.43 | 15.18 |
| 2458702.89287 | 14.58 | 0.02 | MASTER-OAFA | 14.43 | 14.92 | 14.91 | 14.83 | 14.90 | 14.63 | 15.41 | 15.14 |
| 2458702.89543 | 14.55 | 0.02 | MASTER-OAFA | 14.44 | 14.93 | 14.88 | 14.85 | 14.90 | 14.60 | 15.44 | 15.13 |
| 2458704.71152 | 14.65 | 0.02 | MASTER-IAC | 14.42 | 14.90 | 14.87 | 14.83 | 14.92 | 14.63 | 15.47 | 15.16 |
| 2458704.72034 | 14.63 | 0.02 | MASTER-IAC | 14.42 | 14.90 | 14.89 | 14.85 | 14.91 | 14.61 | 15.43 | 15.18 |
| 2458704.86405 | 14.66 | 0.02 | MASTER-OAFA | 14.43 | 14.94 | 14.87 | 14.83 | 14.88 | 14.62 | 15.48 | 15.16 |
| 2458704.88777 | 14.68 | 0.01 | MASTER-OAFA | 14.45 | 14.91 | 14.86 | 14.85 | 14.89 | 14.61 | 15.42 | 15.18 |
| 2458709.51876 | 14.66 | 0.03 | MASTER-Kislovodsk | 14.44 | 14.96 | 14.84 | 14.80 | 14.92 | 14.62 | 15.45 | 15.15 |
| 2458709.52376 | 14.67 | 0.03 | MASTER-Kislovodsk | 14.44 | 14.96 | 14.85 | 14.80 | 14.84 | 14.62 | 15.45 | 15.23 |
| 2458710.68296 | 14.84 | 0.04 | MASTER-IAC | 14.41 | 14.90 | 14.91 | 14.89 | 14.87 | 14.58 | 15.40 | 15.25 |
| 2458710.68296 | 14.86 | 0.02 | MASTER-IAC | 14.44 | 14.91 | 14.92 | 14.81 | 14.89 | 14.61 | 15.47 | 15.15 |
| 2458710.70493 | 14.87 | 0.02 | MASTER-IAC | 14.46 | 14.93 | 14.89 | 14.78 | 14.89 | 14.61 | 15.42 | 15.18 |
| 2458710.70493 | 14.91 | 0.02 | MASTER-IAC | 14.45 | 14.93 | 14.89 | 14.82 | 14.85 | 14.62 | 15.41 | 15.21 |
| 2458711.74343 | 15.00 | 0.04 | MASTER-IAC | 14.42 | 14.98 | 14.80 | 14.83 | 14.86 | 14.65 | 15.46 | 15.19 |
| 2458711.85438 | 14.97 | 0.07 | MASTER-OAFA | 14.43 | 14.99 | 14.91 | 14.81 | 14.76 | 14.62 | 15.40 | 15.32 |
| 2458713.34824 | 15.12 | 0.05 | MASTER-Tunka | 14.45 | 14.81 | 14.84 | 14.81 | 14.90 | 14.68 | 15.51 | 15.20 |
| 2458713.35429 | 15.10 | 0.06 | MASTER-Tunka | 14.38 | 14.85 | 14.89 | 14.88 | 14.88 | 14.75 | 15.38 | 15.17 |
| 2458714.50407 | 15.22 | 0.02 | MASTER-Kislovodsk | 14.42 | 14.90 | 14.90 | 14.86 | 14.91 | 14.62 | 15.44 | 15.14 |
| 2458714.50804 | 15.22 | 0.03 | MASTER-Kislovodsk | 14.43 | 14.86 | 14.92 | 14.86 | 14.87 | 14.62 | 15.43 | 15.20 |
| 2458714.67332 | 15.21 | 0.01 | MASTER-IAC | 14.43 | 14.94 | 14.88 | 14.80 | 14.89 | 14.64 | 15.43 | 15.17 |
| 2458714.67333 | 15.17 | 0.03 | MASTER-IAC | 14.45 | 14.91 | 14.91 | 14.84 | 14.85 | 14.61 | 15.48 | 15.12 |
| 2458714.68138 | 15.15 | 0.03 | MASTER-IAC | 14.45 | 14.95 | 14.86 | 14.81 | 14.81 | 14.65 | 15.44 | 15.20 |
| 2458714.68138 | 15.26 | 0.03 | MASTER-IAC | 14.42 | 14.91 | 14.86 | 14.84 | 14.83 | 14.66 | 15.44 | 15.24 |
| 2458859.57782 | 15.02 | 0.04 | MASTER-OAFA | 14.44 | 14.89 | 14.88 | 14.84 | 14.95 | 14.62 | 15.46 | 15.08 |
| 2458859.60191 | 15.05 | 0.03 | MASTER-OAFA | 14.43 | 14.89 | 14.85 | 14.84 | 14.94 | 14.63 | 15.46 | 15.13 |
| 2458940.22886 | 14.47 | 0.02 | MASTER-SAAO | 14.40 | 14.95 | 14.90 | 14.86 | 14.87 | 14.61 | 15.43 | 15.16 |
| 2458940.22886 | 14.50 | 0.01 | MASTER-SAAO | 14.42 | 14.92 | 14.88 | 14.85 | 14.85 | 14.64 | 15.45 | 15.17 |
| 2458940.22979 | 14.50 | 0.02 | MASTER-SAAO | 14.45 | 14.91 | 14.86 | 14.81 | 14.91 | 14.61 | 15.45 | 15.19 |
| 2458940.22979 | 14.47 | 0.01 | MASTER-SAAO | 14.43 | 14.94 | 14.89 | 14.84 | 14.89 | 14.59 | 15.44 | 15.17 |
| 2458940.23071 | 14.49 | 0.01 | MASTER-SAAO | 14.46 | 14.91 | 14.88 | 14.85 | 14.87 | 14.62 | 15.43 | 15.17 |
| 2458940.23071 | 14.45 | 0.02 | MASTER-SAAO | 14.42 | 14.91 | 14.87 | 14.84 | 14.91 | 14.63 | 15.45 | 15.16 |
| 2458940.23163 | 14.48 | 0.02 | MASTER-SAAO | 14.44 | 14.95 | 14.89 | 14.83 | 14.85 | 14.61 | 15.45 | 15.17 |
| 2458940.23163 | 14.49 | 0.02 | MASTER-SAAO | 14.41 | 14.92 | 14.89 | 14.85 | 14.90 | 14.59 | 15.44 | 15.18 |
| 2458940.23256 | 14.49 | 0.01 | MASTER-SAAO | 14.43 | 14.92 | 14.89 | 14.85 | 14.89 | 14.62 | 15.42 | 15.18 |
| 2458940.23256 | 14.49 | 0.02 | MASTER-SAAO | 14.42 | 14.91 | 14.88 | 14.84 | 14.93 | 14.59 | 15.43 | 15.18 |
| 2458940.23348 | 14.47 | 0.02 | MASTER-SAAO | 14.42 | 14.94 | 14.89 | 14.84 | 14.92 | 14.57 | 15.43 | 15.18 |
| 2458940.23348 | 14.48 | 0.01 | MASTER-SAAO | 14.42 | 14.92 | 14.88 | 14.83 | 14.90 | 14.62 | 15.42 | 15.19 |
| 2458940.23440 | 14.51 | 0.02 | MASTER-SAAO | 14.43 | 14.92 | 14.89 | 14.85 | 14.92 | 14.59 | 15.44 | 15.17 |
| 2458940.23440 | 14.49 | 0.02 | MASTER-SAAO | 14.42 | 14.92 | 14.90 | 14.85 | 14.85 | 14.58 | 15.45 | 15.18 |
| 2458940.23532 | 14.48 | 0.01 | MASTER-SAAO | 14.41 | 14.91 | 14.88 | 14.85 | 14.88 | 14.60 | 15.44 | 15.18 |
| 2458940.23532 | 14.48 | 0.01 | MASTER-SAAO | 14.42 | 14.92 | 14.89 | 14.85 | 14.87 | 14.59 | 15.44 | 15.18 |
| 2458940.23624 | 14.50 | 0.02 | MASTER-SAAO | 14.44 | 14.92 | 14.88 | 14.84 | 14.90 | 14.58 | 15.45 | 15.18 |
| 2458940.23624 | 14.45 | 0.01 | MASTER-SAAO | 14.42 | 14.93 | 14.89 | 14.83 | 14.87 | 14.61 | 15.44 | 15.18 |
| 2458940.23717 | 14.47 | 0.01 | MASTER-SAAO | 14.42 | 14.90 | 14.87 | 14.85 | 14.89 | 14.61 | 15.44 | 15.18 |
| 2458940.23717 | 14.47 | 0.01 | MASTER-SAAO | 14.43 | 14.92 | 14.89 | 14.85 | 14.90 | 14.59 | 15.44 | 15.17 |
| 2458940.23809 | 14.49 | 0.01 | MASTER-SAAO | 14.44 | 14.92 | 14.89 | 14.85 | 14.89 | 14.60 | 15.42 | 15.18 |
| 2458940.23809 | 14.48 | 0.02 | MASTER-SAAO | 14.42 | 14.92 | 14.89 | 14.85 | 14.85 | 14.62 | 15.45 | 15.17 |
| 2458940.23901 | 14.48 | 0.01 | MASTER-SAAO | 14.42 | 14.90 | 14.87 | 14.84 | 14.89 | 14.60 | 15.44 | 15.19 |
| 2458940.23901 | 14.49 | 0.01 | MASTER-SAAO | 14.43 | 14.94 | 14.89 | 14.83 | 14.88 | 14.61 | 15.43 | 15.18 |
| 2458940.23994 | 14.49 | 0.01 | MASTER-SAAO | 14.44 | 14.93 | 14.89 | 14.85 | 14.89 | 14.62 | 15.43 | 15.16 |
| 2458940.23994 | 14.49 | 0.02 | MASTER-SAAO | 14.43 | 14.89 | 14.87 | 14.86 | 14.91 | 14.60 | 15.45 | 15.17 |
| 2458940.24086 | 14.48 | 0.02 | MASTER-SAAO | 14.42 | 14.91 | 14.89 | 14.86 | 14.86 | 14.61 | 15.43 | 15.18 |
| 2458940.24086 | 14.49 | 0.01 | MASTER-SAAO | 14.42 | 14.93 | 14.88 | 14.83 | 14.89 | 14.63 | 15.44 | 15.17 |
| 2458940.24179 | 14.48 | 0.01 | MASTER-SAAO | 14.42 | 14.91 | 14.88 | 14.84 | 14.91 | 14.60 | 15.43 | 15.19 |
| 2458940.24179 | 14.48 | 0.01 | MASTER-SAAO | 14.42 | 14.93 | 14.88 | 14.84 | 14.88 | 14.62 | 15.44 | 15.17 |
| 2458940.24271 | 14.50 | 0.01 | MASTER-SAAO | 14.42 | 14.93 | 14.90 | 14.85 | 14.87 | 14.60 | 15.44 | 15.17 |
| 2458940.24271 | 14.50 | 0.01 | MASTER-SAAO | 14.42 | 14.92 | 14.88 | 14.84 | 14.88 | 14.61 | 15.44 | 15.18 |
| 2458940.24363 | 14.47 | 0.01 | MASTER-SAAO | 14.41 | 14.93 | 14.89 | 14.85 | 14.88 | 14.62 | 15.42 | 15.17 |
| 2458940.24363 | 14.47 | 0.02 | MASTER-SAAO | 14.44 | 14.91 | 14.88 | 14.85 | 14.89 | 14.58 | 15.44 | 15.18 |
| 2458940.24455 | 14.49 | 0.02 | MASTER-SAAO | 14.41 | 14.92 | 14.91 | 14.86 | 14.91 | 14.60 | 15.42 | 15.17 |
| 2458940.24455 | 14.48 | 0.01 | MASTER-SAAO | 14.44 | 14.90 | 14.87 | 14.85 | 14.89 | 14.61 | 15.45 | 15.17 |
| 2458940.24548 | 14.47 | 0.02 | MASTER-SAAO | 14.41 | 14.93 | 14.90 | 14.85 | 14.88 | 14.61 | 15.44 | 15.16 |
| 2458940.24548 | 14.46 | 0.02 | MASTER-SAAO | 14.42 | 14.94 | 14.89 | 14.84 | 14.90 | 14.60 | 15.44 | 15.16 |
| 2458940.24640 | 14.47 | 0.01 | MASTER-SAAO | 14.42 | 14.92 | 14.89 | 14.85 | 14.87 | 14.61 | 15.45 | 15.17 |
| 2458940.24917 | 14.49 | 0.01 | MASTER-SAAO | 14.42 | 14.93 | 14.88 | 14.84 | 14.87 | 14.60 | 15.45 | 15.17 |
| 2458940.25010 | 14.48 | 0.02 | MASTER-SAAO | 14.42 | 14.92 | 14.89 | 14.85 | 14.88 | 14.60 | 15.47 | 15.15 |
| 2458940.25010 | 14.46 | 0.01 | MASTER-SAAO | 14.42 | 14.92 | 14.89 | 14.84 | 14.90 | 14.60 | 15.43 | 15.19 |
| 2458940.25102 | 14.49 | 0.02 | MASTER-SAAO | 14.44 | 14.92 | 14.88 | 14.84 | 14.92 | 14.61 | 15.44 | 15.17 |
| 2458940.25102 | 14.48 | 0.01 | MASTER-SAAO | 14.42 | 14.93 | 14.90 | 14.84 | 14.89 | 14.60 | 15.45 | 15.16 |
| 2458940.25194 | 14.49 | 0.01 | MASTER-SAAO | 14.42 | 14.92 | 14.88 | 14.85 | 14.87 | 14.62 | 15.44 | 15.17 |
| 2458940.25194 | 14.47 | 0.01 | MASTER-SAAO | 14.44 | 14.91 | 14.87 | 14.83 | 14.88 | 14.63 | 15.45 | 15.18 |
| 2458940.25287 | 14.48 | 0.01 | MASTER-SAAO | 14.43 | 14.93 | 14.90 | 14.84 | 14.88 | 14.61 | 15.43 | 15.18 |
| 2458940.25287 | 14.48 | 0.01 | MASTER-SAAO | 14.44 | 14.90 | 14.88 | 14.85 | 14.88 | 14.60 | 15.44 | 15.18 |
| 2458940.25380 | 14.48 | 0.02 | MASTER-SAAO | 14.42 | 14.94 | 14.92 | 14.86 | 14.85 | 14.61 | 15.43 | 15.17 |
| 2458940.25380 | 14.48 | 0.01 | MASTER-SAAO | 14.44 | 14.92 | 14.88 | 14.84 | 14.90 | 14.62 | 15.44 | 15.17 |
| 2458940.25472 | 14.50 | 0.01 | MASTER-SAAO | 14.42 | 14.93 | 14.89 | 14.84 | 14.88 | 14.62 | 15.43 | 15.17 |
| 2458940.25472 | 14.46 | 0.02 | MASTER-SAAO | 14.42 | 14.93 | 14.90 | 14.84 | 14.90 | 14.58 | 15.45 | 15.17 |
| 2458940.25564 | 14.49 | 0.01 | MASTER-SAAO | 14.42 | 14.92 | 14.88 | 14.85 | 14.88 | 14.60 | 15.42 | 15.19 |
| 2458940.25564 | 14.47 | 0.01 | MASTER-SAAO | 14.43 | 14.93 | 14.88 | 14.83 | 14.91 | 14.62 | 15.44 | 15.17 |
| 2458940.25656 | 14.48 | 0.01 | MASTER-SAAO | 14.42 | 14.92 | 14.88 | 14.84 | 14.91 | 14.62 | 15.43 | 15.17 |
| 2458940.25656 | 14.48 | 0.01 | MASTER-SAAO | 14.43 | 14.93 | 14.90 | 14.84 | 14.88 | 14.59 | 15.45 | 15.17 |
| 2458940.25748 | 14.47 | 0.01 | MASTER-SAAO | 14.42 | 14.91 | 14.89 | 14.85 | 14.88 | 14.62 | 15.42 | 15.18 |
| 2458940.25748 | 14.47 | 0.01 | MASTER-SAAO | 14.43 | 14.93 | 14.88 | 14.84 | 14.87 | 14.60 | 15.45 | 15.17 |
| 2458940.25840 | 14.48 | 0.02 | MASTER-SAAO | 14.42 | 14.94 | 14.91 | 14.85 | 14.87 | 14.62 | 15.42 | 15.17 |
| 2458940.25841 | 14.47 | 0.01 | MASTER-SAAO | 14.44 | 14.92 | 14.89 | 14.85 | 14.88 | 14.60 | 15.44 | 15.17 |
| 2458940.25933 | 14.47 | 0.01 | MASTER-SAAO | 14.41 | 14.92 | 14.88 | 14.84 | 14.88 | 14.61 | 15.43 | 15.18 |
| 2458940.25933 | 14.45 | 0.01 | MASTER-SAAO | 14.43 | 14.92 | 14.88 | 14.84 | 14.89 | 14.61 | 15.46 | 15.17 |
| 2458940.26025 | 14.48 | 0.01 | MASTER-SAAO | 14.43 | 14.92 | 14.88 | 14.84 | 14.87 | 14.59 | 15.45 | 15.18 |
| 2458940.26025 | 14.47 | 0.01 | MASTER-SAAO | 14.42 | 14.94 | 14.90 | 14.84 | 14.87 | 14.62 | 15.44 | 15.17 |
| 2458940.26118 | 14.48 | 0.02 | MASTER-SAAO | 14.42 | 14.93 | 14.89 | 14.83 | 14.92 | 14.61 | 15.45 | 15.16 |
| 2458940.26118 | 14.48 | 0.01 | MASTER-SAAO | 14.45 | 14.93 | 14.88 | 14.84 | 14.88 | 14.62 | 15.45 | 15.17 |
| 2458940.26210 | 14.48 | 0.02 | MASTER-SAAO | 14.43 | 14.93 | 14.91 | 14.85 | 14.90 | 14.59 | 15.43 | 15.17 |
| 2458940.26210 | 14.45 | 0.02 | MASTER-SAAO | 14.40 | 14.89 | 14.87 | 14.85 | 14.90 | 14.64 | 15.44 | 15.16 |
| 2458940.26302 | 14.49 | 0.01 | MASTER-SAAO | 14.43 | 14.92 | 14.89 | 14.85 | 14.88 | 14.61 | 15.42 | 15.19 |
| 2458940.26302 | 14.47 | 0.01 | MASTER-SAAO | 14.41 | 14.91 | 14.87 | 14.84 | 14.88 | 14.62 | 15.44 | 15.18 |
| 2458940.26394 | 14.47 | 0.01 | MASTER-SAAO | 14.41 | 14.92 | 14.89 | 14.84 | 14.88 | 14.62 | 15.42 | 15.18 |
| 2458940.26394 | 14.45 | 0.01 | MASTER-SAAO | 14.42 | 14.92 | 14.88 | 14.85 | 14.86 | 14.62 | 15.44 | 15.17 |
| 2458940.26486 | 14.48 | 0.01 | MASTER-SAAO | 14.44 | 14.94 | 14.89 | 14.82 | 14.90 | 14.62 | 15.42 | 15.18 |
| 2458940.26486 | 14.44 | 0.01 | MASTER-SAAO | 14.43 | 14.92 | 14.89 | 14.85 | 14.87 | 14.61 | 15.43 | 15.18 |
| 2458940.26579 | 14.48 | 0.01 | MASTER-SAAO | 14.43 | 14.94 | 14.89 | 14.83 | 14.89 | 14.61 | 15.43 | 15.18 |
| 2458940.26579 | 14.45 | 0.01 | MASTER-SAAO | 14.44 | 14.92 | 14.88 | 14.83 | 14.90 | 14.63 | 15.44 | 15.17 |
| 2458940.26671 | 14.48 | 0.01 | MASTER-SAAO | 14.44 | 14.90 | 14.88 | 14.86 | 14.88 | 14.63 | 15.43 | 15.16 |
| 2458940.26671 | 14.48 | 0.01 | MASTER-SAAO | 14.44 | 14.93 | 14.90 | 14.85 | 14.90 | 14.60 | 15.43 | 15.17 |
| 2458940.26763 | 14.48 | 0.01 | MASTER-SAAO | 14.42 | 14.93 | 14.89 | 14.85 | 14.87 | 14.61 | 15.43 | 15.17 |
| 2458940.26763 | 14.48 | 0.01 | MASTER-SAAO | 14.44 | 14.92 | 14.88 | 14.84 | 14.88 | 14.61 | 15.45 | 15.17 |
| 2458940.26856 | 14.46 | 0.01 | MASTER-SAAO | 14.42 | 14.94 | 14.90 | 14.84 | 14.87 | 14.60 | 15.44 | 15.18 |
| 2458940.26856 | 14.47 | 0.01 | MASTER-SAAO | 14.45 | 14.94 | 14.89 | 14.83 | 14.88 | 14.62 | 15.44 | 15.18 |
| 2458940.26948 | 14.47 | 0.01 | MASTER-SAAO | 14.42 | 14.92 | 14.88 | 14.84 | 14.91 | 14.61 | 15.43 | 15.17 |
| 2458940.26948 | 14.48 | 0.02 | MASTER-SAAO | 14.44 | 14.89 | 14.86 | 14.82 | 14.91 | 14.64 | 15.45 | 15.18 |
| 2458940.27040 | 14.48 | 0.01 | MASTER-SAAO | 14.42 | 14.90 | 14.87 | 14.85 | 14.87 | 14.62 | 15.45 | 15.18 |
| 2458940.27040 | 14.48 | 0.01 | MASTER-SAAO | 14.43 | 14.91 | 14.88 | 14.84 | 14.86 | 14.60 | 15.45 | 15.19 |
| 2458940.27133 | 14.46 | 0.02 | MASTER-SAAO | 14.43 | 14.93 | 14.89 | 14.84 | 14.85 | 14.60 | 15.45 | 15.18 |
| 2458940.27133 | 14.48 | 0.01 | MASTER-SAAO | 14.42 | 14.93 | 14.88 | 14.83 | 14.89 | 14.60 | 15.46 | 15.17 |
| 2458940.27225 | 14.48 | 0.01 | MASTER-SAAO | 14.44 | 14.93 | 14.89 | 14.84 | 14.89 | 14.59 | 15.44 | 15.18 |
| 2458940.27225 | 14.47 | 0.01 | MASTER-SAAO | 14.42 | 14.92 | 14.88 | 14.83 | 14.89 | 14.61 | 15.45 | 15.17 |
| 2458940.27317 | 14.46 | 0.01 | MASTER-SAAO | 14.42 | 14.93 | 14.89 | 14.84 | 14.87 | 14.62 | 15.42 | 15.18 |
| 2458940.27317 | 14.47 | 0.02 | MASTER-SAAO | 14.46 | 14.95 | 14.89 | 14.83 | 14.90 | 14.60 | 15.44 | 15.17 |
| 2458940.27410 | 14.48 | 0.02 | MASTER-SAAO | 14.44 | 14.93 | 14.90 | 14.86 | 14.88 | 14.59 | 15.44 | 15.16 |
| 2458940.27410 | 14.46 | 0.01 | MASTER-SAAO | 14.44 | 14.92 | 14.89 | 14.85 | 14.86 | 14.63 | 15.44 | 15.16 |
| 2458940.27502 | 14.47 | 0.01 | MASTER-SAAO | 14.42 | 14.94 | 14.89 | 14.84 | 14.88 | 14.62 | 15.42 | 15.18 |
| 2458940.27502 | 14.44 | 0.01 | MASTER-SAAO | 14.42 | 14.93 | 14.88 | 14.83 | 14.87 | 14.59 | 15.44 | 15.19 |
| 2458940.27594 | 14.45 | 0.02 | MASTER-SAAO | 14.43 | 14.93 | 14.90 | 14.85 | 14.89 | 14.57 | 15.46 | 15.17 |
| 2458940.27594 | 14.47 | 0.01 | MASTER-SAAO | 14.43 | 14.91 | 14.87 | 14.83 | 14.89 | 14.61 | 15.46 | 15.17 |
| 2458940.27687 | 14.47 | 0.02 | MASTER-SAAO | 14.41 | 14.91 | 14.88 | 14.85 | 14.89 | 14.63 | 15.41 | 15.18 |
| 2458940.27687 | 14.45 | 0.01 | MASTER-SAAO | 14.43 | 14.93 | 14.87 | 14.82 | 14.90 | 14.60 | 15.44 | 15.19 |
| 2458940.27780 | 14.48 | 0.02 | MASTER-SAAO | 14.41 | 14.94 | 14.89 | 14.83 | 14.90 | 14.63 | 15.41 | 15.18 |
| 2458940.27780 | 14.45 | 0.02 | MASTER-SAAO | 14.40 | 14.93 | 14.88 | 14.83 | 14.86 | 14.64 | 15.43 | 15.18 |
| 2458940.27872 | 14.48 | 0.01 | MASTER-SAAO | 14.44 | 14.92 | 14.88 | 14.83 | 14.89 | 14.61 | 15.44 | 15.18 |
| 2458940.27872 | 14.46 | 0.01 | MASTER-SAAO | 14.44 | 14.91 | 14.87 | 14.84 | 14.87 | 14.60 | 15.45 | 15.18 |
| 2458940.27964 | 14.47 | 0.01 | MASTER-SAAO | 14.43 | 14.92 | 14.89 | 14.85 | 14.91 | 14.61 | 15.43 | 15.17 |
| 2458940.27964 | 14.46 | 0.01 | MASTER-SAAO | 14.43 | 14.93 | 14.89 | 14.85 | 14.88 | 14.60 | 15.42 | 15.18 |
| 2458940.28056 | 14.46 | 0.01 | MASTER-SAAO | 14.44 | 14.90 | 14.87 | 14.85 | 14.90 | 14.63 | 15.43 | 15.17 |
| 2458940.28056 | 14.47 | 0.02 | MASTER-SAAO | 14.44 | 14.91 | 14.86 | 14.83 | 14.90 | 14.64 | 15.47 | 15.16 |
| 2458940.28150 | 14.48 | 0.02 | MASTER-SAAO | 14.44 | 14.96 | 14.89 | 14.83 | 14.86 | 14.62 | 15.44 | 15.17 |
| 2458940.28150 | 14.45 | 0.01 | MASTER-SAAO | 14.43 | 14.93 | 14.89 | 14.84 | 14.87 | 14.59 | 15.47 | 15.17 |
| 2458940.28242 | 14.46 | 0.01 | MASTER-SAAO | 14.43 | 14.94 | 14.90 | 14.84 | 14.89 | 14.62 | 15.43 | 15.16 |
| 2458940.28242 | 14.47 | 0.01 | MASTER-SAAO | 14.46 | 14.92 | 14.88 | 14.83 | 14.88 | 14.59 | 15.45 | 15.19 |
| 2458940.28334 | 14.48 | 0.02 | MASTER-SAAO | 14.44 | 14.92 | 14.89 | 14.85 | 14.93 | 14.59 | 15.44 | 15.17 |
| 2458940.28334 | 14.48 | 0.01 | MASTER-SAAO | 14.43 | 14.95 | 14.90 | 14.83 | 14.87 | 14.62 | 15.42 | 15.18 |
| 2458940.28427 | 14.48 | 0.02 | MASTER-SAAO | 14.44 | 14.94 | 14.91 | 14.84 | 14.91 | 14.60 | 15.42 | 15.17 |
| 2458940.28427 | 14.47 | 0.01 | MASTER-SAAO | 14.42 | 14.92 | 14.87 | 14.83 | 14.88 | 14.62 | 15.43 | 15.19 |
| 2458940.28519 | 14.48 | 0.02 | MASTER-SAAO | 14.42 | 14.94 | 14.91 | 14.85 | 14.92 | 14.62 | 15.41 | 15.16 |
| 2458940.28519 | 14.49 | 0.02 | MASTER-SAAO | 14.43 | 14.95 | 14.89 | 14.83 | 14.83 | 14.63 | 15.43 | 15.18 |
| 2458940.28611 | 14.44 | 0.02 | MASTER-SAAO | 14.41 | 14.94 | 14.91 | 14.85 | 14.90 | 14.60 | 15.42 | 15.17 |
| 2458940.28611 | 14.47 | 0.01 | MASTER-SAAO | 14.44 | 14.93 | 14.89 | 14.84 | 14.89 | 14.60 | 15.42 | 15.18 |
| 2458940.28704 | 14.46 | 0.02 | MASTER-SAAO | 14.45 | 14.93 | 14.89 | 14.84 | 14.90 | 14.58 | 15.45 | 15.18 |
| 2458940.28704 | 14.45 | 0.01 | MASTER-SAAO | 14.43 | 14.93 | 14.88 | 14.83 | 14.87 | 14.63 | 15.43 | 15.19 |
| 2458940.28796 | 14.47 | 0.03 | MASTER-SAAO | 14.43 | 14.96 | 14.93 | 14.85 | 14.86 | 14.58 | 15.42 | 15.18 |
| 2458940.28796 | 14.48 | 0.01 | MASTER-SAAO | 14.43 | 14.95 | 14.89 | 14.82 | 14.88 | 14.61 | 15.43 | 15.19 |
| 2458940.28888 | 14.48 | 0.01 | MASTER-SAAO | 14.43 | 14.90 | 14.87 | 14.84 | 14.87 | 14.65 | 15.43 | 15.18 |
| 2458940.28888 | 14.49 | 0.01 | MASTER-SAAO | 14.44 | 14.92 | 14.87 | 14.83 | 14.86 | 14.62 | 15.44 | 15.19 |
| 2458940.28981 | 14.48 | 0.02 | MASTER-SAAO | 14.44 | 14.96 | 14.90 | 14.84 | 14.85 | 14.60 | 15.43 | 15.17 |
| 2458940.28981 | 14.48 | 0.02 | MASTER-SAAO | 14.43 | 14.95 | 14.89 | 14.82 | 14.87 | 14.60 | 15.43 | 15.20 |

| | | | | | | | | | | |
|---|---|---|---|---|---|---|---|---|---|---|
| 2458940.29073 | 14.48 | 0.02 | MASTER-SAAO | 14.41 | 14.93 | 14.90 | 14.84 | 14.86 | 14.60 | 15.44 | 15.18 |
| 2458940.29073 | 14.46 | 0.01 | MASTER-SAAO | 14.43 | 14.93 | 14.89 | 14.83 | 14.88 | 14.61 | 15.42 | 15.19 |
| 2458940.29165 | 14.48 | 0.01 | MASTER-SAAO | 14.42 | 14.93 | 14.88 | 14.83 | 14.85 | 14.62 | 15.45 | 15.18 |
| 2458940.29165 | 14.48 | 0.02 | MASTER-SAAO | 14.46 | 14.92 | 14.88 | 14.83 | 14.92 | 14.63 | 15.42 | 15.17 |
| 2458940.29257 | 14.49 | 0.01 | MASTER-SAAO | 14.43 | 14.93 | 14.89 | 14.84 | 14.88 | 14.64 | 15.43 | 15.17 |
| 2458940.29257 | 14.47 | 0.01 | MASTER-SAAO | 14.45 | 14.91 | 14.86 | 14.82 | 14.88 | 14.64 | 15.43 | 15.20 |
| 2458940.29350 | 14.46 | 0.02 | MASTER-SAAO | 14.43 | 14.90 | 14.86 | 14.84 | 14.93 | 14.65 | 15.43 | 15.16 |
| 2458940.29350 | 14.45 | 0.02 | MASTER-SAAO | 14.46 | 14.95 | 14.89 | 14.83 | 14.86 | 14.64 | 15.44 | 15.16 |
| 2458940.29442 | 14.49 | 0.01 | MASTER-SAAO | 14.45 | 14.93 | 14.89 | 14.84 | 14.91 | 14.63 | 15.43 | 15.16 |
| 2458940.29442 | 14.48 | 0.02 | MASTER-SAAO | 14.46 | 14.92 | 14.88 | 14.83 | 14.90 | 14.65 | 15.43 | 15.17 |
| 2458940.29535 | 14.51 | 0.02 | MASTER-SAAO | 14.44 | 14.91 | 14.85 | 14.82 | 14.87 | 14.65 | 15.44 | 15.19 |
| 2458940.29535 | 14.50 | 0.02 | MASTER-SAAO | 14.44 | 14.89 | 14.87 | 14.85 | 14.87 | 14.66 | 15.44 | 15.17 |
| 2458940.29627 | 14.45 | 0.02 | MASTER-SAAO | 14.40 | 14.94 | 14.89 | 14.83 | 14.91 | 14.61 | 15.45 | 15.16 |
| 2458940.29627 | 14.45 | 0.03 | MASTER-SAAO | 14.45 | 14.93 | 14.89 | 14.84 | 14.80 | 14.64 | 15.43 | 15.20 |
| 2458940.29719 | 14.45 | 0.02 | MASTER-SAAO | 14.41 | 14.92 | 14.90 | 14.86 | 14.88 | 14.61 | 15.43 | 15.17 |
| 2458940.29719 | 14.47 | 0.02 | MASTER-SAAO | 14.43 | 14.92 | 14.88 | 14.83 | 14.83 | 14.64 | 15.45 | 15.19 |
| 2458940.29812 | 14.45 | 0.02 | MASTER-SAAO | 14.41 | 14.92 | 14.88 | 14.84 | 14.91 | 14.65 | 15.42 | 15.16 |
| 2458940.29812 | 14.53 | 0.03 | MASTER-SAAO | 14.42 | 14.89 | 14.82 | 14.81 | 14.86 | 14.67 | 15.45 | 15.20 |
| 2458940.29904 | 14.48 | 0.02 | MASTER-SAAO | 14.46 | 14.93 | 14.90 | 14.85 | 14.89 | 14.61 | 15.44 | 15.16 |
| 2458940.29904 | 14.46 | 0.02 | MASTER-SAAO | 14.41 | 14.91 | 14.88 | 14.86 | 14.86 | 14.64 | 15.43 | 15.17 |
| 2458940.29996 | 14.52 | 0.02 | MASTER-SAAO | 14.44 | 14.92 | 14.89 | 14.85 | 14.88 | 14.66 | 15.44 | 15.15 |
| 2458940.29996 | 14.48 | 0.02 | MASTER-SAAO | 14.43 | 14.92 | 14.88 | 14.85 | 14.85 | 14.60 | 15.42 | 15.20 |
| 2458940.30089 | 14.48 | 0.01 | MASTER-SAAO | 14.42 | 14.90 | 14.87 | 14.85 | 14.89 | 14.62 | 15.43 | 15.18 |
| 2458940.30089 | 14.46 | 0.03 | MASTER-SAAO | 14.42 | 14.99 | 14.91 | 14.82 | 14.86 | 14.63 | 15.39 | 15.19 |
| 2458940.30181 | 14.50 | 0.02 | MASTER-SAAO | 14.44 | 14.93 | 14.88 | 14.83 | 14.91 | 14.59 | 15.44 | 15.18 |
| 2458940.30181 | 14.44 | 0.03 | MASTER-SAAO | 14.43 | 14.97 | 14.93 | 14.83 | 14.85 | 14.64 | 15.41 | 15.17 |
| 2458940.30273 | 14.47 | 0.01 | MASTER-SAAO | 14.42 | 14.93 | 14.88 | 14.82 | 14.89 | 14.62 | 15.43 | 15.19 |
| 2458940.30273 | 14.47 | 0.02 | MASTER-SAAO | 14.41 | 14.95 | 14.91 | 14.83 | 14.86 | 14.65 | 15.43 | 15.16 |
| 2458940.30365 | 14.48 | 0.03 | MASTER-SAAO | 14.44 | 14.89 | 14.84 | 14.82 | 14.94 | 14.63 | 15.42 | 15.19 |
| 2458940.30365 | 14.44 | 0.01 | MASTER-SAAO | 14.43 | 14.92 | 14.87 | 14.83 | 14.87 | 14.64 | 15.44 | 15.18 |
| 2458940.32226 | 14.47 | 0.01 | MASTER-SAAO | 14.44 | 14.91 | 14.87 | 14.82 | 14.89 | 14.65 | 15.45 | 15.18 |
| 2458940.32318 | 14.49 | 0.03 | MASTER-SAAO | 14.44 | 14.92 | 14.84 | 14.79 | 14.87 | 14.65 | 15.47 | 15.19 |
| 2458940.32410 | 14.47 | 0.02 | MASTER-SAAO | 14.45 | 14.91 | 14.86 | 14.82 | 14.92 | 14.63 | 15.44 | 15.19 |
| 2458940.32503 | 14.49 | 0.04 | MASTER-SAAO | 14.48 | 14.92 | 14.85 | 14.78 | 14.88 | 14.70 | 15.43 | 15.20 |
| 2458940.32595 | 14.48 | 0.03 | MASTER-SAAO | 14.43 | 14.90 | 14.85 | 14.81 | 14.90 | 14.67 | 15.47 | 15.17 |
| 2458940.32687 | 14.44 | 0.04 | MASTER-SAAO | 14.43 | 14.89 | 14.83 | 14.82 | 14.84 | 14.65 | 15.41 | 15.24 |
| 2458940.32779 | 14.49 | 0.04 | MASTER-SAAO | 14.48 | 14.92 | 14.83 | 14.79 | 14.90 | 14.69 | 15.44 | 15.18 |
| 2458940.32871 | 14.45 | 0.03 | MASTER-SAAO | 14.42 | 14.93 | 14.87 | 14.82 | 14.80 | 14.64 | 15.47 | 15.19 |
| 2458940.32963 | 14.48 | 0.03 | MASTER-SAAO | 14.46 | 14.91 | 14.84 | 14.80 | 14.94 | 14.65 | 15.44 | 15.18 |
| 2458940.33056 | 14.47 | 0.03 | MASTER-SAAO | 14.45 | 14.89 | 14.85 | 14.81 | 14.86 | 14.65 | 15.45 | 15.21 |
| 2458940.33148 | 14.45 | 0.02 | MASTER-SAAO | 14.44 | 14.92 | 14.87 | 14.82 | 14.87 | 14.67 | 15.45 | 15.17 |
| 2458940.33240 | 14.45 | 0.03 | MASTER-SAAO | 14.41 | 14.93 | 14.85 | 14.79 | 14.89 | 14.60 | 15.48 | 15.19 |
| 2458940.33333 | 14.48 | 0.02 | MASTER-SAAO | 14.42 | 14.91 | 14.84 | 14.81 | 14.89 | 14.64 | 15.46 | 15.18 |
| 2458940.33425 | 14.42 | 0.03 | MASTER-SAAO | 14.40 | 14.91 | 14.84 | 14.81 | 14.86 | 14.65 | 15.44 | 15.20 |
| 2458940.33518 | 14.51 | 0.03 | MASTER-SAAO | 14.48 | 14.89 | 14.86 | 14.83 | 14.93 | 14.60 | 15.45 | 15.19 |
| 2458940.33610 | 14.47 | 0.03 | MASTER-SAAO | 14.44 | 14.93 | 14.86 | 14.78 | 14.89 | 14.68 | 15.42 | 15.21 |
| 2458940.33702 | 14.45 | 0.02 | MASTER-SAAO | 14.41 | 14.92 | 14.88 | 14.85 | 14.85 | 14.64 | 15.42 | 15.18 |
| 2458940.33795 | 14.41 | 0.04 | MASTER-SAAO | 14.47 | 14.96 | 14.87 | 14.78 | 14.82 | 14.67 | 15.47 | 15.19 |
| 2458940.33887 | 14.46 | 0.04 | MASTER-SAAO | 14.46 | 14.86 | 14.84 | 14.82 | 14.87 | 14.68 | 15.41 | 15.22 |
| 2458940.33979 | 14.42 | 0.03 | MASTER-SAAO | 14.42 | 14.92 | 14.84 | 14.80 | 14.93 | 14.62 | 15.41 | 15.22 |
| 2458940.34072 | 14.44 | 0.03 | MASTER-SAAO | 14.47 | 14.92 | 14.85 | 14.79 | 14.87 | 14.65 | 15.49 | 15.18 |
| 2458940.34164 | 14.50 | 0.03 | MASTER-SAAO | 14.48 | 14.92 | 14.85 | 14.78 | 14.93 | 14.64 | 15.47 | 15.18 |
| 2458940.34257 | 14.47 | 0.05 | MASTER-SAAO | 14.43 | 14.92 | 14.81 | 14.77 | 14.84 | 14.66 | 15.52 | 15.18 |
| 2458940.34349 | 14.49 | 0.04 | MASTER-SAAO | 14.45 | 14.96 | 14.88 | 14.79 | 14.82 | 14.64 | 15.50 | 15.17 |
| 2458940.34441 | 14.44 | 0.04 | MASTER-SAAO | 14.51 | 14.91 | 14.86 | 14.82 | 14.84 | 14.60 | 15.46 | 15.21 |
| 2458940.34533 | 14.44 | 0.04 | MASTER-SAAO | 14.45 | 14.90 | 14.84 | 14.77 | 14.93 | 14.68 | 15.48 | 15.18 |
| 2458940.34626 | 14.42 | 0.06 | MASTER-SAAO | 14.42 | 14.89 | 14.76 | 14.74 | 14.86 | 14.65 | 15.51 | 15.23 |
| 2458940.34718 | 14.45 | 0.05 | MASTER-SAAO | 14.41 | 14.92 | 14.80 | 14.76 | 14.89 | 14.70 | 15.44 | 15.22 |
| 2458940.34811 | 14.44 | 0.04 | MASTER-SAAO | 14.47 | 14.93 | 14.85 | 14.76 | 14.91 | 14.67 | 15.51 | 15.16 |
| 2458940.34903 | 14.50 | 0.04 | MASTER-SAAO | 14.44 | 14.94 | 14.85 | 14.75 | 14.83 | 14.65 | 15.49 | 15.22 |
| 2458940.34995 | 14.44 | 0.08 | MASTER-SAAO | 14.45 | 14.84 | 14.73 | 14.77 | 14.92 | 14.70 | 15.51 | 15.19 |
| 2458940.35088 | 14.47 | 0.04 | MASTER-SAAO | 14.47 | 14.91 | 14.85 | 14.78 | 14.87 | 14.67 | 15.49 | 15.18 |
| 2458940.35180 | 14.49 | 0.07 | MASTER-SAAO | 14.50 | 14.89 | 14.77 | 14.76 | 14.82 | 14.66 | 15.52 | 15.22 |